# Classification testing: A new framework for drawing qualitative conclusions from quantitative estimates*

Andrew C. Eggers†
and
Zikai Li‡

August 23, 2026

**Abstract**

Social scientists rely on hypothesis testing to support their research conclusions, but the standard tests are designed for testing one hypothesis rather than adjudicating between rival possibilities. We develop a new framework, "classification testing", as an alternative. Instead of selecting one hypothesis to test, a researcher conducting a classification test decides what qualitative distinctions ("classes") are most substantively relevant; the test either assigns the estimand to a class with error control similar to that of a conventional hypothesis test, or declares the result inconclusive. We argue that classification testing is superior to current practice not just when the objective is to adjudicate between rival possibilities but also when there is one research hypothesis to be tested, because classification testing exposes that hypothesis to refutation. We illustrate the framework by applying it to a well-known media experiment and offer an R package to aid in implementation.

*We thank Macartan Humphreys, Charlotte Cavaillé, Ben Lauderdale, and participants at the 2026 Chicago Area Methods Seminar (CHARMS) for helpful comments. We used Claude Code (Opus 4.6-4.8) as a research and coding assistant in preparing this manuscript. Any remaining errors are our own.

†Professor, Department of Political Science, University of Chicago

‡PhD Student, Department of Political Science, University of Chicago

# 1 Introduction

Much of quantitative social science research is concerned with estimation (Lundberg et al. 2021), but often the main conclusions of that research are communicated in qualitative terms: an effect is present, or is positive/negative, or is negligible (Edlin and Love 2022). These qualitative judgments are typically supported by a hypothesis test (usually a two-tailed test; occasionally a one-tailed test or an equivalence test), and therefore come with frequentist error guarantees about the probability of falsely making the stated claim. In social science, despite critiques of how hypothesis testing is conducted and interpreted (Gigerenzer et al. 2004; Gill 1999; McShane et al. 2019; Nickerson 2000; e.g., Rosnow and Rosenthal 1992), it remains the standard way to justify conclusions (Emmert-Streib 2024; Witteloostuijn and Hugten 2022).

In this paper we offer a new approach to generating principled research conclusions that we argue is superior to existing hypothesis testing practice. One major problem with current practice is that it treats rival possibilities asymmetrically: it is possible to reject the null hypothesis (and thus affirm a complementary research hypothesis with error control) but not to reject the research hypothesis (and affirm the null). For example, following common guidance (Gerber and Green 2012; Hales 2024), a researcher may use a one-tailed test with the null hypothesis $H_0 : \theta \leq 0$ to test the research hypothesis that an estimand $\theta$ is positive. If the null is rejected, the researcher can make an error-controlled declaration that the estimand is positive; if it is not rejected, no error-controlled conclusion is possible. In effect, the research hypothesis can be "proven" right, but not wrong. Current practice is especially unsatisfying when the goal is not to test a specific hypothesis (such as "$\theta$ is positive") but instead to answer an open question such as "is $\theta$ positive or negative?": which possibility should be the null hypothesis and therefore vulnerable to falsification?

As an improvement, we propose *classification testing*. Like standard hypothesis tests, classification tests allow researchers to make error-controlled qualitative statements about quantitative estimands. The difference is that a standard test can affirm only one desig-nated hypothesis (when the complementary null hypothesis is rejected), whereas a classification test can affirm any of its classes with the same error control. Instead of selecting a single research hypothesis for testing (as in current practice), a researcher conducting a classification test begins by partitioning the parameter space into two or more substantively relevant and logically exhaustive classes. (For example: "Is the estimand positive or is it (weakly) negative?" "Is the estimand negligible, substantial and positive, or substantial and negative?") The researcher may, but is not required to, designate one class as their own hypothesis. After estimation, the researcher either makes an error-controlled conclusion in favor of one of the classes or, if the data do not point strongly enough to one class, declares the result "inconclusive". Although the framework is general, we focus on three useful and simple cases: a sign classification test that aims to determine the sign of an estimand, a magnitude classification test that distinguishes between negligible and substantial values of the estimand, and a sign-and-magnitude test that combines both objectives.

Classification testing stands to help research practice in both direct and indirect ways. The benefit compared to current hypothesis testing practice is most apparent when the purpose of the research is to adjudicate among rival possibilities: in such circumstances, researchers who switch to classification testing can simply describe those possibilities in a pre-registration rather than artificially designate one as the research hypothesis. When the objective is to test a single hypothesis, switching to classification testing means putting the research hypothesis at risk of refutation (not just risk of non-affirmation), making the test more informative. Because classification testing does not require researchers to choose a single hypothesis and more readily produces error-controlled affirmative conclusions, switching to classification testing may have the more indirect benefit of encouraging researchers to pursue genuinely open questions and uncover more unexpected findings.

Our classification test framework relates most closely to two areas of previous research. First, our motivation and method relate to equivalence testing, a form of hypothesis testing that replaces the standard hypothesis that an effect is present with the hypothesis that the effect is small in magnitude (Hartman and Hidalgo 2018; Lakens 2017a; Rainey 2014; Schuirmann 1987). Like an equivalence test, our magnitude classification test allows an error-controlled conclusion of negligibility, but it is more powerful than the standard equivalence testing procedure (TOST) and it also allows an error-controlled affirmation of non-negligibility. Second, our framework builds on and unifies previously proposed tests that allow for more than one error-controlled conclusion. It can be seen as an instance of the partitioning principle proposed by Finner and Strassburger (2002). Our sign classification test mirrors Kaiser (1960)'s "directional two-sided test" and a later proposal by Jones and Tukey (2000; see also Rice and Krakauer 2023). It is equivalent to running two one-sided tests without the need for multiplicity correction.[1] Like our sign-and-magnitude classification test, Goeman et al. (2010)'s "three-sided hypothesis testing" (see also Isager and Fitzgerald 2024; Riffenburgh and Wang 2025) allows a conclusion of equivalence and (signed) non-equivalence.[2] Our general framework yields these and other tests and is simple and intuitive enough to appeal to applied researchers.

In the next section, we explain the need for error-controlled qualitative statements in quantitative social science and why current practice is unsatisfactory for that purpose (Section 2). We then provide an intuitive overview of the classification testing approach (Section 3). We next introduce the general framework for classification tests (Section 4), present three practically useful classification test types (Section 5), discuss how multiple classification tests can be combined (Section 6), and illustrate our approach by applying it to a recently published field experiment (Section 7). Section 8 discusses the practical implementation of classification tests. Section 9 concludes.

[1]The boundary point (typically zero) should be on the same side in both tests. We explain this further in Section 5.1.

[2]As we show, our approach is more powerful.

# 2 Motivation

## 2.1 The Value of Qualitative Conclusions about Quantitative Estimands

Perhaps surprisingly, most quantitative studies in social science omit numbers both from their headline findings (Edlin and Love 2022) and (in our experience) from their reviews of prior literature. Instead, the results are discussed in *qualitative* terms: the target quantity is not zero (e.g., something "matters"), or is positive or negative in sign, or is negligible or substantial in magnitude. These qualitative statements are often supported by hypothesis testing.

Several observers have criticized this tendency to convert quantitative findings into qualitative conclusions, often as part of a critique of prevalent hypothesis testing practices. Cumming (2014) and McShane et al. (2019) argue, for example, that social scientists should focus less on the results of hypothesis tests and more on point estimates, standard errors, and confidence intervals in assessing what we learn from a piece of research. Similarly, Edlin and Love (2022) urges a move from a "null hypothesis testing culture" to an "estimation culture".

While we certainly welcome a more continuous form of reasoning about evidence, we believe the conversion of estimates into qualitative conclusions is inevitable to some extent, and even desirable in some cases. Often the theoretically relevant characteristic of an estimand is its sign or whether it is zero, in which case the precise magnitude is of secondary importance.[3] When the estimand is relevant for a policy decision, the salient question might be whether a treatment works better than an alternative, and the best simple summary might be a direct, yes-or-no answer rather than an estimate of a treatment effect. In many areas of applied research, the magnitude of focal estimates is difficult to interpret without substantial knowledge of the underlying measures and design.[4] In such cases, the full research paper must provide the context for understanding the estimand and estimate, but it is appropriate and necessary for headline findings to provide a coarsened, qualitative summary, indicating, for example, whether the target estimand is positive or "large" by some pre-determined standard.

[3] For examples, see Ashworth et al. (2021).

## 2.2 The Importance of Error Control

In current social science practice, key qualitative conclusions are typically supported by hypothesis tests. Although we will argue that current hypothesis testing practice has important limitations, we endorse the principle of frequentist error control behind it and use it in classification testing, so here we offer a brief review.

Consider a one-tailed test of the null hypothesis $H_0 : \theta \leq 0$ (where $\theta$ represents an ATE), given an estimate $\widehat{\theta}$ and associated standard error $\sigma$. Suppose we reject the null hypothesis when the estimate is above a point $c$. Assuming $\widehat{\theta} \sim \mathcal{N}(\theta, \sigma^2)$, the probability of rejecting the null is $\Pr(\widehat{\theta} > c) = 1 - \Phi(\frac{c-\theta}{\sigma})$. Of course, this probability is unknown, because $\theta$ is unknown. But we can choose $c$ so that, for all $\theta$, the probability of incorrectly rejecting the null is no higher than a chosen value $\alpha$. For a given $c$, the probability of rejecting the null increases in $\theta$; therefore it is sufficient to choose $c = \Phi^{-1}(1 - \alpha)\,\sigma \approx 1.645\sigma$. We then say that the procedure, and the decisions it produces, are "error-controlled".

The error this guarantee controls is an incorrect affirmation of $H_1$, the complement of the null hypothesis and often the research hypothesis. Take $H_1 : \theta > 0$, the complement of the null $H_0 : \theta \leq 0$. Because the two are complements, $H_0$ is true for exactly the values of $\theta$ for which $H_1$ is false, so the Type I error rate $\Pr(\text{reject } H_0 \mid H_0 \text{ true})$ is equal to $\Pr(\text{reject } H_0 \mid H_1 \text{ false})$. When $\widehat{\theta}$ is in the rejection region we reject $H_0$, which is equiavlent to affirming $H_1$ because $H_1$ is true if and only if $H_0$ is false. We can thus interpret the Type

[4] Standardizing units helps, but interpretation remains difficult without additional information about variation in the relevant measures.

I error rate as the probability of affirming $H_1$ when $H_1$ is false, and error control holds this probability at or below $\alpha$.

Frequentist error control offers reassurance about a procedure's long-run performance, but without further information it does not limit the probability that a specific result is incorrect. Whatever prior one holds over $\theta$, the *ex ante* probability of a mistaken rejection cannot exceed $\alpha$ (because this is the maximum probability of an erroneous rejection across values of $\theta$). Yet the probability that a *given* rejection is mistaken depends on that prior and can be arbitrarily high. If $\theta$ is known to be zero (say, the effect of exercising now on last week's health), any rejection is erroneous. Judging a specific result therefore takes more than the error guarantee of the procedure.

## 2.3 Why Current Practice Is Inadequate

Current hypothesis testing practice makes error-controlled qualitative statements against a pre-specified null hypothesis, and therefore in favor of a research hypothesis standing in contrast to that null hypothesis. While this is a valuable function, we argue that current practice is inadequate — both as a general way to produce principled qualitative statements and as a technique for testing a research hypothesis — because current practice reserves error-controlled affirmation for the research hypothesis alone.

By "current hypothesis testing practice", we mean Wald-style null hypothesis significance tests as used by applied researchers in political science, economics, sociology, and related fields. The default testing procedure (implemented by all common regression software) is a two-tailed test at significance level $\alpha = .05$ of the null hypothesis $H_0 : \theta = 0$. If the estimate's magnitude is at least 1.96 times the standard error, the researcher rejects the null and declares the result "significant"; otherwise the researcher reports a failure to reject the null (a "null result").[5] When the research focuses on the sign of an estimand (as it often

[5] Gigerenzer et al. (2004) call this procedure and associated interpretations the "null ritual".

does[6]), Gerber and Green (2012) and Lakens (2017b) instead advocate a (pre-registered) one-tailed test, which more readily allows an error-controlled affirmation that the estimand is in the hypothesized direction.[7] Finally, when the researcher seeks to make a principled conclusion that an estimand is *small*, Rainey (2014), Lakens (2017a), and Hartman and Hidalgo (2018) advocate an equivalence test, which tests the null that the estimand is *not* small ($H_0 : |\theta| > \delta$ for some $\delta$). If the researcher can reject that null (typically using the two one-sided tests procedure, or TOST[8]), they can make an error-controlled statement in favor of the estimand's negligibility.

A key feature of these testing procedures is that they treat the null and the research hypotheses asymmetrically: only the null hypothesis is tested, and rejecting it licenses an error-controlled statement in favor of its complement, the research hypothesis. Thus a standard two-tailed test can license a claim that the estimand is not zero; a one-tailed test can license a claim that the estimand is positive or negative, depending on the sign of the null; an equivalence test can license a claim that the estimand is small. Notably, there is no provision for affirming the null hypothesis with error control.

A common justification for this asymmetry is that the null hypothesis represents a default view that is provisionally retained unless the researcher can mount sufficiently strong evidence against it; only then can she declare victory for her preferred alternative.[9] This justification clearly doesn't apply when there are two or more genuinely rival views and the researcher is not arguing for one of them. Standard practice nonetheless requires the researcher to designate one view as the null hypothesis and assess whether it can be rejected.

This justification is also unsatisfying even when the test pits the researcher's hypothesis against a null that represents a prevailing view. In any of these tests, a failure to reject the null hypothesis can occur either because the data is much more consistent with that null than with the researcher's hypothesis or because the data does not point strongly in either direction. So even if we claim to "provisionally retain" the null hypothesis when it is not rejected, the correct evidential interpretation of a failure to reject the null (absent other information) is to view the question as not resolved. The test thus delivers either an error-controlled verdict that "the researcher's hypothesis is correct (the null is incorrect)" or a determination that "the question is unresolved". An error-controlled rejection of the researcher's hypothesis is impossible.

[6]For example, Ofosu and Posner (2023) report that 90% of pre-analysis plans in political science and economics include a directional hypothesis.

[7]Our survey of recent experimental papers in top political science journals (2020-2024) indicates that this advice is followed, but only to a limited extent: of 49 cases where a PAP includes a directional hypothesis, 9 (18%) include a one-tailed test.

[8]Equivalently, one can check if the 90% confidence interval lies entirely in the interval $(-\delta, \delta)$.

[9]For example, Hartman and Hidalgo (2018) argue for an equivalence testing approach to placebo tests on these grounds; similarly, Rainey (2014)'s title recommends equivalence testing when "Arguing for a negligible effect".

In our view, the asymmetry also generates bad incentives for researchers in choosing and conducting research projects. In a system where affirmative findings are rewarded and only research hypotheses can be affirmed, self-interested researchers would choose a research hypothesis that is *ex ante* highly likely and take pains to obtain findings consistent with it. Our journals may therefore have too many studies of essentially settled questions instead of genuinely open ones, and too many tests (and affirmations) of conventional wisdom instead of minority viewpoints.

# 3 Classification Testing in Outline

Our framework ("classification testing") replaces current hypothesis testing practice with a procedure designed to adjudicate between two or more possibilities. A researcher conducting a classification test partitions $\theta$ into "classes" that correspond to distinct qualitative conclusions about $\theta$. The specific partition depends on what distinctions are relevant in a particular domain (which requires substantive context and researcher judgment) and, when possible, should be pre-registered. The researcher also chooses a significance level $\alpha$ (e.g., .05) that indicates the maximum allowable probability of assigning $\theta$ to the wrong class. Given an estimate $\hat{\theta}$ and a standard error $\hat{\sigma}$ (and assuming, as is commonly the case in applied work, an asymptotically normal sampling distribution for $\hat{\theta}$), the procedure then either assigns $\theta$ to a class or declares the result inconclusive. An assignment lets the researcher affirmatively claim that $\theta$ belongs to that class with the reassurance that the probability of an incorrect classification is capped at $\alpha$.

The next two sections explain how classification testing optimally maps an estimate and standard error into an error-controlled classification decision. This mapping is subtle for all but the simplest classification problems, but an approximation to the optimal procedure can be described very simply:

1. Choose (and ideally pre-register) a partition of $\theta$ into classes. Choose a significance level $\alpha$.
2. Obtain an estimate $\hat{\theta}$ and standard error $\hat{\sigma}$, and construct a $1 - 2\alpha$ (e.g., 90%, if $\alpha = .05$) confidence interval $\hat{\theta} \pm z_{1-\alpha}\hat{\sigma}$.[10]
3. If the CI lies entirely in one class, report that $\theta$ belongs to that class; otherwise, report "inconclusive".

As we show below, this simplified version perfectly describes our procedure in the case where there are just two classes separated by a single boundary point (the case considered by Kaiser (1960) and Jones and Tukey (2000)). In other problems, the simplified version closely approximates our procedure when the standard error of the estimator is small relative to the size of the classes (in a sense we make precise) but fails to achieve the targeted error rate when the estimator is noisier. Thus "assign $\theta$ to a class if and only if the 90% confidence interval fits in it" provides a useful shorthand for the procedure, but the full implementation is necessary to retain the desired error control for all but the simplest classification problems.

As suggested by the simplified version of our procedure, classification testing is related to the common practice of informally drawing qualitative conclusions about an estimand from a confidence interval — e.g., concluding that an estimand is probably small if the 95% confi-dence interval only includes small values.[11] In essence, classification testing formalizes this practice while achieving desired error control more precisely than is possible with any fixed confidence interval. It thus allows researchers to avoid the shortcomings of current practice while retaining the ability to make qualitative conclusions with optimal error control.

[10]Note $z_{1-\alpha} \equiv \Phi^{-1}(1-\alpha)$, where $\Phi(\cdot)$ is the standard normal CDF.

We now proceed to define the framework in a general manner and show how to implement it in practically relevant cases.

# 4 A Classification Framework for Statistical Inference

## 4.1 Framework Definition

Consider a scalar estimand $\theta$ and an estimator $\hat{\theta}$ with asymptotic distribution $\hat{\theta} \sim N(\theta, \sigma^2)$, where $\sigma^2$ may be unknown but is consistently estimated.

**Partition the parameter space** into $K$ mutually exclusive classes $C_1, C_2, \ldots, C_K$, where $\bigcup_i C_i = \Theta$ and $C_i \cap C_j = \emptyset$ for $i \neq j$. Figure 1 shows three practically useful types of classification:

- $C_- = (-\infty, 0]$, $C_+ = (0, \infty)$ (sign classification)
- $C_0 = (-\delta, \delta)$, $C_\pm = (-\infty, -\delta] \cup [\delta, \infty)$ (magnitude classification)
- $C_0 = (-\delta, \delta)$, $C_- = (-\infty, -\delta]$, $C_+ = [\delta, \infty)$ (sign-and-magnitude classification)

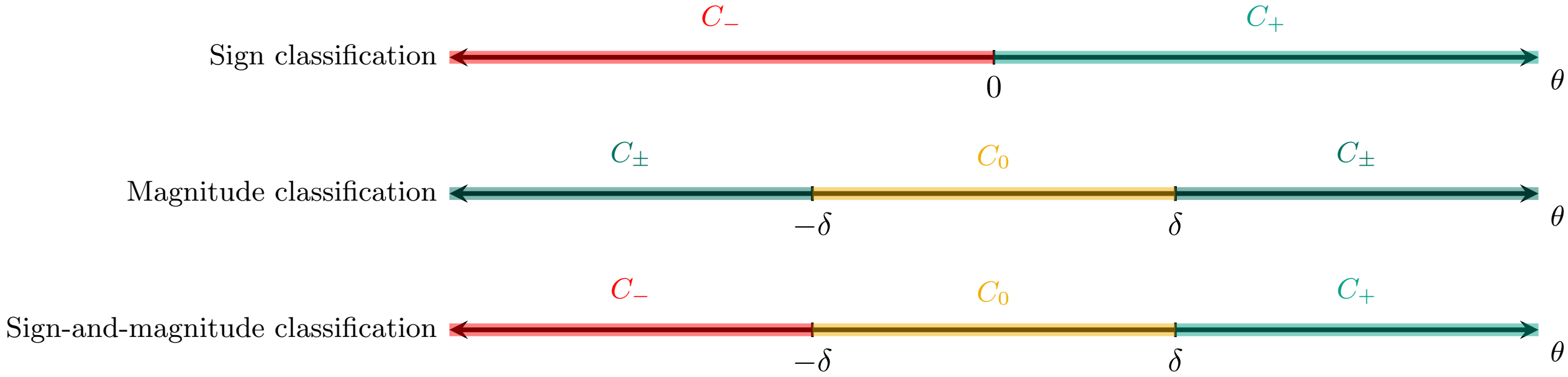

Figure 1: Simple classification problems

Define **acceptance regions** $A_j$ for each class $C_j$, where $\hat{\theta} \in A_j$ leads to classification of

[11] Relatedly, Cumming (2014), McShane et al. (2019), and Greenland et al. (2016) advocate de-emphasizing hypothesis testing and focusing instead on estimates and confidence intervals.

$\theta$ as belonging to class $C_j$. If $\hat{\theta}$ does not fall in any acceptance region then no classification is made: the result is "Inconclusive". Figure 2 depicts possible acceptance regions for the simple classification problems in Figure 1. Our framework can be viewed as a generalization of the Neyman-Pearson framework for hypothesis testing. In that framework, the researcher partitions the parameter space into two classes, a null hypothesis $H_0$ and its complement $H_1$. The acceptance region $A_1$ is chosen so that $\Pr(\hat{\theta} \in A_1 \mid \theta \in H_0) \leq \alpha$: the probability of affirming $H_1$ when it is false is at most $\alpha$. Our framework generalizes this approach to allow for the possibility of (1) affirming the "null" class (which a standard test can reject but cannot affirm), and (2) partitioning $\theta$ into more than two classes.

Classification testing is also an instance of the partitioning principle (Finner and Strassburger 2002). For a partition of the parameter space into disjoint "alternative" hypotheses (equivalent to classes $C_1, C_2, \ldots, C_K$), testing the complementary null for each class (and, on rejection, affirming that class) at level $\alpha$ controls the family-wise error at $\alpha$ without a multiplicity correction. This is because at most one class is true, so at most we can reject the complementary null for one class.

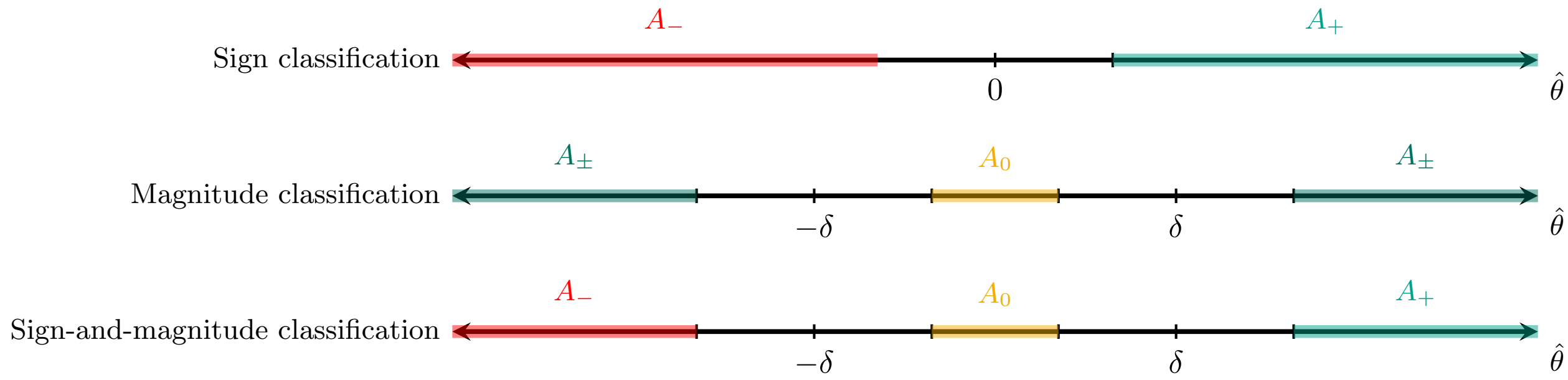

Figure 2: Illustrative acceptance regions for simple classification problems

## 4.2 Error Control

A misclassification occurs when the procedure places the estimand into a class that does not contain $\theta$. Our goal is to construct acceptance regions such that the probability of misclassification is no higher than $\alpha$ for any value of $\theta$. A **size-$\alpha$ classification rule** satisfies:

for all $k$ and all $\theta \in C_k$,

$$P_\theta\left(\hat{\theta} \in \bigcup_{j \neq k} A_j\right) \leq \alpha.$$

That is, the probability that $\theta$ is classified into any wrong class is bounded by $\alpha$ uniformly in $\theta$.

A misclassification is thus a wrong affirmation: to classify $\hat{\theta}$ into $A_j$ is to affirm $\theta \in C_j$, and that affirmation is wrong when $\theta \notin C_j$. Under the size-$\alpha$ condition, the affirmation of any class is error-controlled: whatever the true class, the probability of affirming a class that does not contain $\theta$ is at most $\alpha$.

## 4.3 Worst-Case Power and the Maximin Criterion

We seek to make as many positive classifications as possible subject to the size constraint. We therefore choose the size-$\alpha$ classification rule that maximizes what we call "worst-case power."

**Definition 1** (Worst-case power and optimal test). *The* worst-case power *of a classification rule* $(A_1, \dots, A_K)$ *is*

$$W(A_1, \dots, A_K) := \min_j \inf_{\theta \in C_j} P_\theta(\hat{\theta} \in A_j),$$

*i.e., the smallest probability of a correct classification over all classes and all* $\theta$ *in each class. A size-*$\alpha$ *classification rule is* optimal *if it attains the maximum of* $W$ *among all size-*$\alpha$ *rules:*

$$\max_{(A_1, \dots, A_K): \textit{size} \leq \alpha} W(A_1, \dots, A_K)$$

Our optimality criterion is a "maximin" one (Lehmann and Romano 2022, Ch. 8): among size-$\alpha$ rules, maximize the smallest probability of correctly declaring a class (i.e., worst-case power). For a single-boundary partition (as in sign testing), the test that satisifies the maximin criterion is also a uniformly most powerful test because the normal family has monotone likelihood ratio in $\hat{\theta}$ (Lehmann and Romano 2022, Ch. 3). With two or more boundaries no uniformly most powerful test exists, since no single rule is most powerful for every class at once. We therefore adopt the maximin criterion for the general case.

For the practically relevant two- and three-class cases we focus on, this criterion and the size constraint jointly imply that acceptance regions must be drawn so that, at every class boundary, the misclassification probability approaches $\alpha$ from each adjacent class.[12] In more complex cases, it may be impossible to achieve the rate $\alpha$ simultaneously at all boundary points, in which case more elaborate prioritization rules are necessary.

# 5 Three Simple Classification Tests

We now describe three classification tests in order of increasing complexity.

## 5.1 Sign Classification

Applied researchers very often seek to determine the sign of an estimand $\theta$. For example, as many as 90% of pre-analysis plans in political science and economics state a directional hypothesis (Ofosu and Posner 2023), and empirical tests of comparative statics from formal theory tend to focus on the sign of an estimand (Ashworth et al. 2021). In other cases (e.g., Burde et al. 2026), researchers seek to determine whether an estimand is above some threshold other than zero.

In a classification testing approach to these problems, the researcher partitions $\theta$ into two classes separated by a single boundary point and seeks to determine which class $\theta$ belongs to. When the boundary point is at zero, so that $C_{-} = (-\infty, 0]$ and $C_{+} = (0, \infty)$, we have a *sign classification test*. When the boundary is some other point $c$, we can treat the problem as sign classification by redefining the estimand as $\theta - c$. The outcome of the test is an error-controlled conclusion that $\theta$ is positive, an error-controlled conclusion that $\theta$ is negative,[13] or an inconclusive result.

[12] We prove this for the sign-and-magnitude test in Proposition 5 (Appendix Section D.5); the sign and magnitude cases are analogous.

In this case attention focuses on the boundary point at $\theta = 0$, where the misclassification rate will be at a maximum. We have (arbitrarily) assigned this point to $C_-$ ("negative"), so misclassification error occurs when we obtain a large enough estimate that we misclassify it as $C_+$ ("positive"). To ensure this error rate does not exceed $\alpha$, we classify estimates as "positive" if and only if they are in the upper $\alpha$ tail (i.e., at or above the $1-\alpha$ quantile) of the sampling distribution when $\theta = 0$, i.e., if $\hat{\theta} \geq z_{1-\alpha}\sigma$. The same argument applies just on the other side of the boundary (i.e., at the limit as $\theta$ goes to zero from above). By symmetry, the acceptance regions are $A_- = (-\infty, -z_{1-\alpha}\sigma]$ and $A_+ = [z_{1-\alpha}\sigma, \infty)$. Figure 3 illustrates.

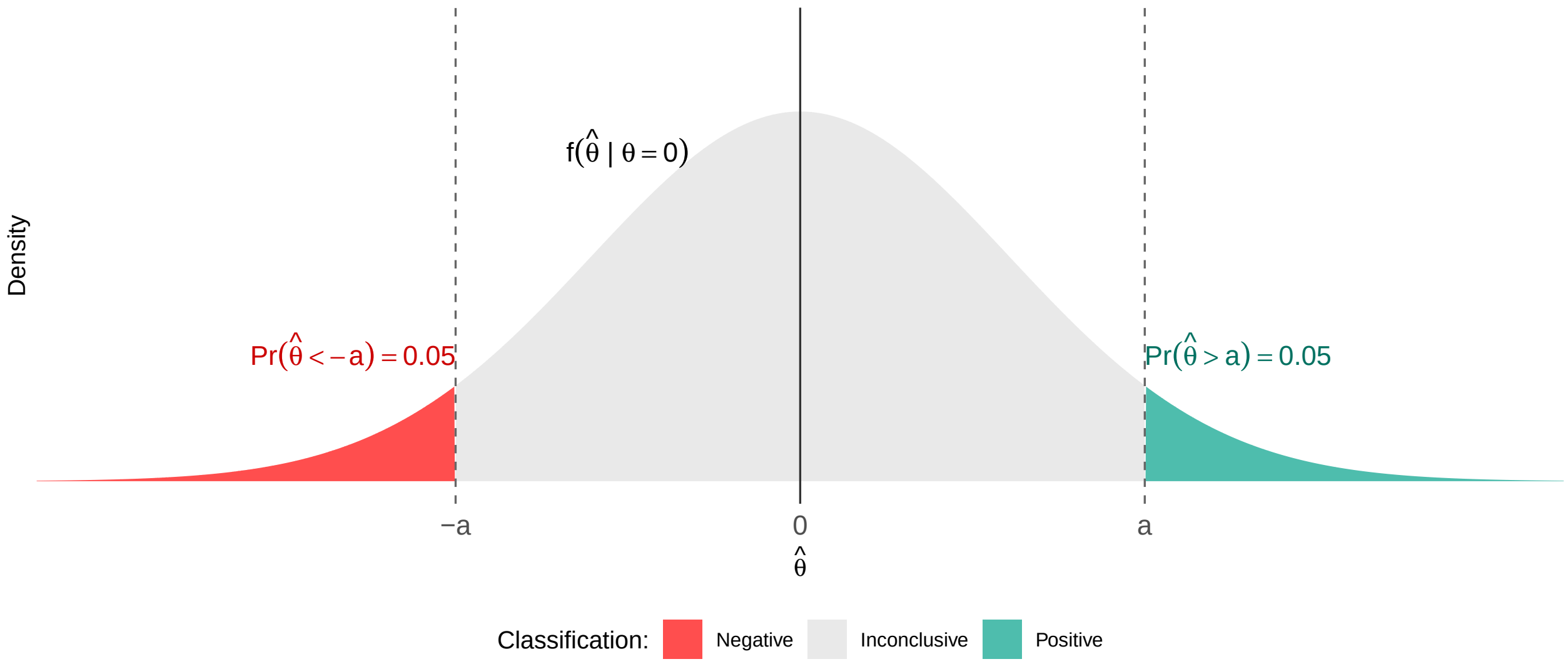

Figure 3: Classifications at $\theta = 0$ for the sign classification test. The inner boundary of the "positive" acceptance region ($a$) is set so that, when $\theta$ is at the upper boundary of the "negative" class (here, $0$), the probability of an incorrect classification (i.e., $\hat{\theta} \geq a$) is exactly $\alpha$ (here, 0.05). The same logic applies to the "negative" acceptance region, producing a symmetric boundary at $-a$. An "inconclusive" result occurs when $\hat{\theta}$ falls in the interval $(-a, a)$.

Sign classification can be seen as running two one-sided null hypothesis significance tests, one where $H_0 : \theta \leq 0$ and the other where $H_0 : \theta > 0$. Suppose a one-sided test has

[13]Technically, "weakly negative".

$H_0 : \theta \leq 0$ and $H_1 : \theta > 0$. In both current practice and our classification procedure, we would reject $H_0$ (declaring that $\theta$ is positive) if $\hat{\theta} > z_{1-\alpha}\sigma$. In current practice, however, any estimate $\hat{\theta} < z_{1-\alpha}\sigma$ produces the same qualitative finding: we are unable to reject the null.[14] But if we set up a second test with $H_0 : \theta > 0$, that test is "free:" we can conduct it without multiplicity correction, because the two tests are mutually exclusive and rejecting one precludes rejecting the other.[15] Our sign classification test exploits this fact to provide a more informative result and permit the researcher to distinguish between very negative estimates (which allow a warranted conclusion that $\theta$ is negative) and intermediate estimates that do not allow us to confidently classify the sign of $\theta$.

Sign classification can thus be implemented simply by assigning $\theta$ to a class (positive or negative) if and only if the $1 - 2\alpha$ confidence interval lands entirely in that class. To see why this works, note that if $\theta = 0$ (so that the correct class is "(weakly) negative"), it is most difficult to classify $\theta$. In this case, a proportion $\alpha$ of $1 - 2\alpha$ CIs will land entirely to the right of 0 (so that $\theta$ will be misclassified as "positive" with frequency $\alpha$). A similar point can be made for $\theta$ at the lower limit of the "positive" class. As we will see, this approach works only as an approximation for the other classification problems we consider.

Although the sign test is uncommon in current practice, it has been suggested several times before (notably by Kaiser (1960) and Jones and Tukey (2000)) and is justified by the partitioning principle (Finner and Strassburger 2002). It also mirrors Cox et al. (1977)'s pragmatic interpretation of the two-tailed test as a way to divide possible results into an upper critical region, a lower critical region, and a central region where "the data by themselves do not supply clear evidence as to the sign" (p.~51). The sign test accomplishes the same division of results, but (unlike Cox's two-tailed test) achieves the correct error control.

[14] As Lakens (2016) explains, researchers who conduct a one-tailed test of a directional hypothesis and obtain an estimate pointing strongly in the opposite direction are free to describe what they found, but cannot make an error-controlled statement because the results do not confirm the research hypothesis.

[15] This does not hold when $H_0 : \theta \leq 0$ in one test and $H_0 : \theta \geq 0$ in the other. In this case, the two tests are not mutually exclusive and, under $\theta = 0$, the error rate can exceed $\alpha$ if both tests are conducted.

## 5.2 Magnitude Classification

In some cases researchers are interested in determining whether or not an estimand $\theta$ is close to (or equal to) a specific value. Placebo/balance tests provide a salient example: under certain conditions, an estimand $\theta$ should be zero if the core assumptions of the research design are valid but not otherwise (Eggers et al. 2024). Although placebo tests are often conducted using a null hypothesis of $\theta = 0$ (so that rejecting the null hypothesis raises concern about the design), Hartman and Hidalgo (2018) instead advise equivalence tests (Lakens 2017a; Rainey 2014; Schuirmann 1987), where the null hypothesis is $|\theta| > \delta$ for some small $\delta$ (so that rejecting the null allows a researcher to draw an error-controlled conclusion in favor of the soundness of the design).

With the classification testing framework, these problems are straightforwardly handled by partitioning $\theta$ into a bounded, "interior" class and its complement. If the interior class is centered on zero, we can call this a *magnitude classification test* with classes $C_0 = (-\delta, \delta)$ ("negligible" or "small", where $\delta \geq 0$) and $C_\pm = (-\infty, -\delta] \cup [\delta, \infty)$ ("substantial", "non-negligible", "large"). If the interior class is centered on some other point $c$ (e.g., because the substantive question is whether $\theta$ is close to $c$ or not), we can redefine the estimand as $\theta - c$ so that the name fits. The outcome of the test is an error-controlled conclusion that $\theta$ is "negligible", an error-controlled conclusion that $\theta$ is "substantial", or an inconclusive result.

Let $a$ refer to the inner limit of the acceptance region for the upper part of the "substantial" class. (By the symmetry of the problem, $-a$ is the inner limit of the acceptance region for the lower part.) We choose $a$ such that

$$P_{\theta=\delta}\left(\hat{\theta} \in (-\infty, -a] \cup [a, \infty)\right) = \alpha, \tag{1}$$

which caps at $\alpha$ the probability of misclassifying a borderline "negligible" $\theta$ as "substantial".

Let $b$ refer to the upper limit of the acceptance region for the "negligible" class; given the symmetry of the problem, the lower limit is $-b$. We choose $b$ such that

$$P_{\theta=\delta}\left(\hat{\theta} \in (-b, b)\right) = \alpha, \tag{2}$$

which caps at $\alpha$ the probability of misclassifying a borderline "substantial" $\theta$ as "negligible".

To develop intuition for how these acceptance regions are constructed, consider proceeding as if this were two sign classification tests put together: set $a = \delta + z_{1-\alpha}\sigma$ and $b = \delta - z_{1-\alpha}\sigma$; equivalently, declare $\theta$ to belong to a class iff the $1-2\alpha$ CI is located entirely in that class. Note that the results would mirror TOST for the "negligible" class and further allow a "substantial" conclusion.

If $\theta$ is precisely estimated, with $\sigma$ small relative to $\delta$, this $1-2\alpha$ CI-based approach comes close to the optimal solution. As $\hat{\theta}$ gets noisier, the approximation diverges further and further from optimality, for two reasons.

The first divergence occurs because, at moderate levels of precision, the approximation for $b$ (the upper boundary of the "negligible" acceptance region) is overly conservative: the probability of misclassifying a borderline "substantial" case (i.e., when $\theta = \delta$ from above) falls below the nominal rate $\alpha$. The reason is that the lower tail of the sampling distribution for $\hat{\theta}$ (given $\theta = \delta$ from above) overshoots the acceptance region for the "negligible" class, so some of the error budget is wasted. Call $\epsilon_b$ the probability of overshooting in this way. Then if we declare $\theta$ "negligible" when the $1-2\alpha$ CI falls entirely between $-\delta$ and $\delta$, the misclassification rate at $\theta = \delta$ from above is $\alpha - \epsilon_b$. The optimal solution addresses this by setting $b$ larger than the $1-2\alpha$ CI implies (enlarging the "negligible" acceptance region compared to the TOST solution), to an extent that depends on the relationship of $\sigma$ to $\delta$.

This divergence between the optimal approach and the $1-2\alpha$ CI approximation affirms long-standing critiques of the standard TOST approach to equivalence testing (Rainey 2014; Schuirmann 1987). As Berger and Hsu (1996) point out, TOST is conservative at any test precision but noticeably so especially when $\sigma \geq \delta/z_{1-\alpha}$: the TOST rejection region then disappears, so the test cannot produce a finding in favor of negligibility. Berger and Hsu (1996)'s equivalence test addresses this problem, but is not widely used by applied researchers. (See also Romano (2005) and Wellek (2010).) As discussed further in Appendix Section E.2, our plug-in approach is simpler, works for quantities not expressed in standard deviation units, and produces the same results on negligibility unless the sample size is very small. Our magnitude classification test is at least as good as any existing equivalence test at testing for a negligible effect (provided that the estimator is asymptotically normal), but it also allows an error-controlled verdict that an effect is substantial.

The second divergence between the optimal classification test and the $1-2\alpha$ CI approximation occurs because, for even noisier estimators, the approximation for $a$ (the inner bound of the "substantial" acceptance region) is anti-conservative: the probability of misclassifying a borderline "negligible" case (i.e., when $\theta=\delta$) goes above the nominal rate $\alpha$. The reason is that, when $\theta=\delta$, part of the lower tail of the sampling distribution for $\hat{\theta}$ is caught by the lower portion of the acceptance region for the "substantial" class. Call $\epsilon_a$ the probability of obtaining an estimate below $-\delta-z_{1-\alpha}\sigma$ (i.e., obtaining a $1-2\alpha$ CI entirely below $-\delta$) when $\theta=\delta$. If we declare $\theta$ "substantial" when the $1-2\alpha$ CI falls entirely below $-\delta$ or above $\delta$, the misclassification rate at $\theta=\delta$ is $\alpha+\epsilon_a$, where $\epsilon_a$ measures the extent to which using the CI-based approximation is anti-conservative. Our approach addresses this by setting $a$ larger than the CI approximation implies (shrinking the "substantial" acceptance regions), again to an extent that depends on the relationship of $\sigma$ to $\delta$.[16]

Figure 4 plots the misclassification rate against $\sigma/\delta$ (which captures the noisiness of the problem), for our procedure (green), the 90% CI rule (dashed), and the 95% CI rule (dotted). At $\theta=\delta$ from above (left panel) both CI rules are conservative; the 90% CI rule has an error rate noticeably below $\alpha$ once $\sigma/\delta$ exceeds about 1/3. At $\theta=\delta$ (right panel), the 90% CI rule is anti-conservative and the 95% CI rule is conservative; the 90% CI rule's error rate exceeds $\alpha$ noticeably once $\sigma/\delta$ is above about 4/3. Our procedure's error rate is close to $\alpha$ throughout.

[16]As $\sigma/\delta$ goes to infinity, $a$ approaches $\delta+z_{1-\alpha/2}\sigma$; thus the test comes to resemble a two-tailed test of $H_0:\theta=0$ with the addition of an acceptance region for the null hypothesis.

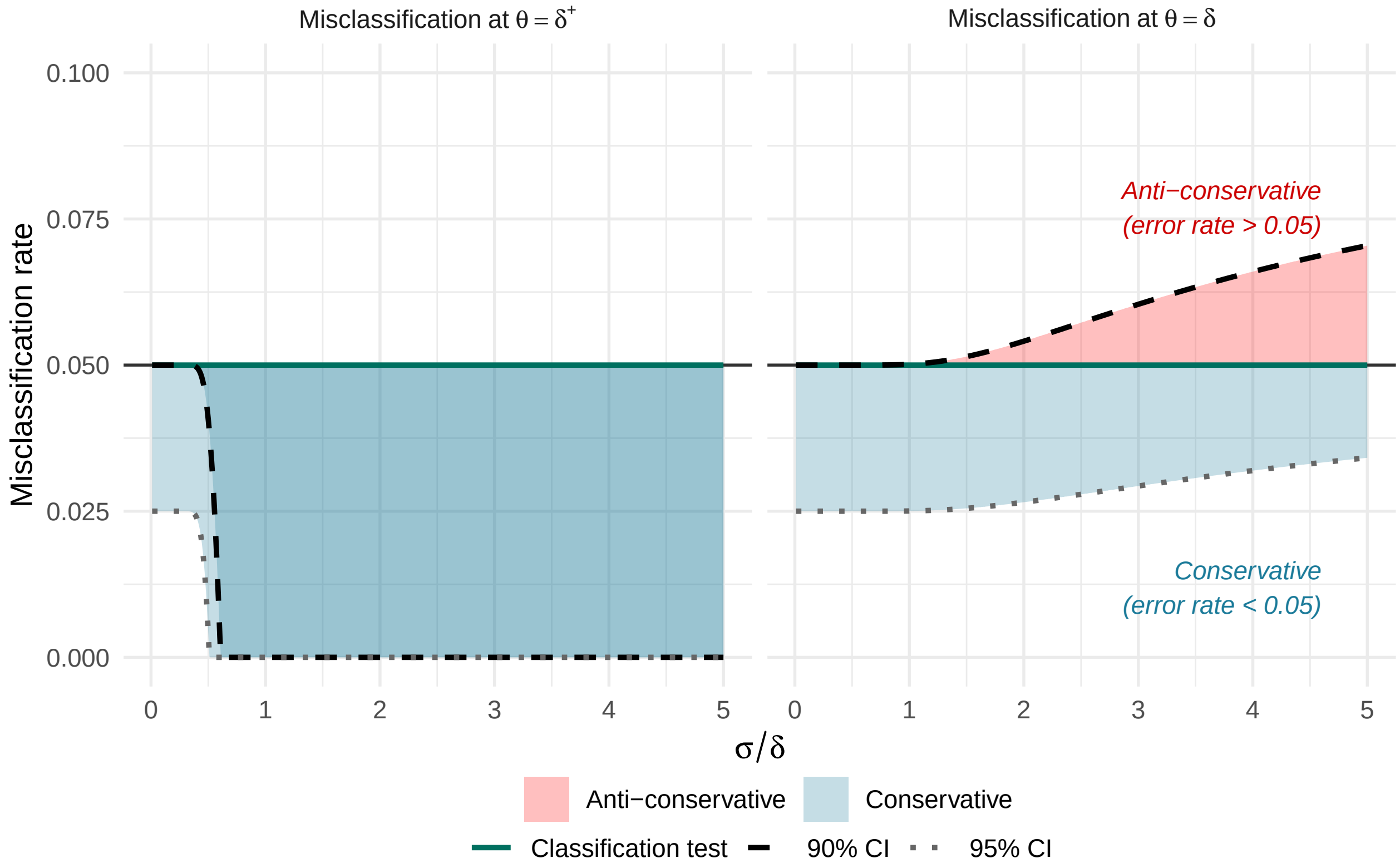

Figure 4: Misclassification rate for $\theta$ just above $\delta$ (left) and at $\delta$ (right) in the magnitude and sign-and-magnitude classification tests using the optimal procedure (solid green line) vs using an approximation based on 90% CIs (dashed line) or 95% CIs (dotted line). Assigning $\theta$ to a class iff the 90% CI lies entirely in that class works well when the estimator is precise but fails to achieve nominal error rates on both sides of $\delta$ when the estimator is noisy.

The left panel of Figure 5 shows how the optimal acceptance regions for a magnitude classification test vary with test precision for fixed $\delta = 1$ and $\alpha = .05$.[17] (Each horizontal cross section gives the acceptance regions for one value of the standard error $\sigma$.) When the estimator is very precise (bottom of the picture), the acceptance regions are about $z_{1-\alpha}\sigma \approx 1.645\sigma$ from the class boundaries, so that as the standard error increases (moving up) the acceptance regions narrow approximately linearly. As the estimator gets noisier, the "negligible" acceptance region narrows less-than-linearly and eventually flares out somewhat. The dotted "W" shape shows the borders of the acceptance regions implied by the rule "assign $\theta$ to a class iff the 90% confidence interval falls entirely in that class"; note that for noisier estimators the optimal acceptance regions are slightly narrower for the "substantial" class and wider for the "negligible" class. The rejection region of the TOST equivalence test lies entirely within the small triangle at the center of the diagram.

[17] Appendix Figure A1 illustrates in detail how acceptance regions are determined for the magnitude test both in a precise estimation case (where the $1-2\alpha$ CI approach closely approximates our solution) and an imprecise estimation case (where the two approaches diverge substantially).

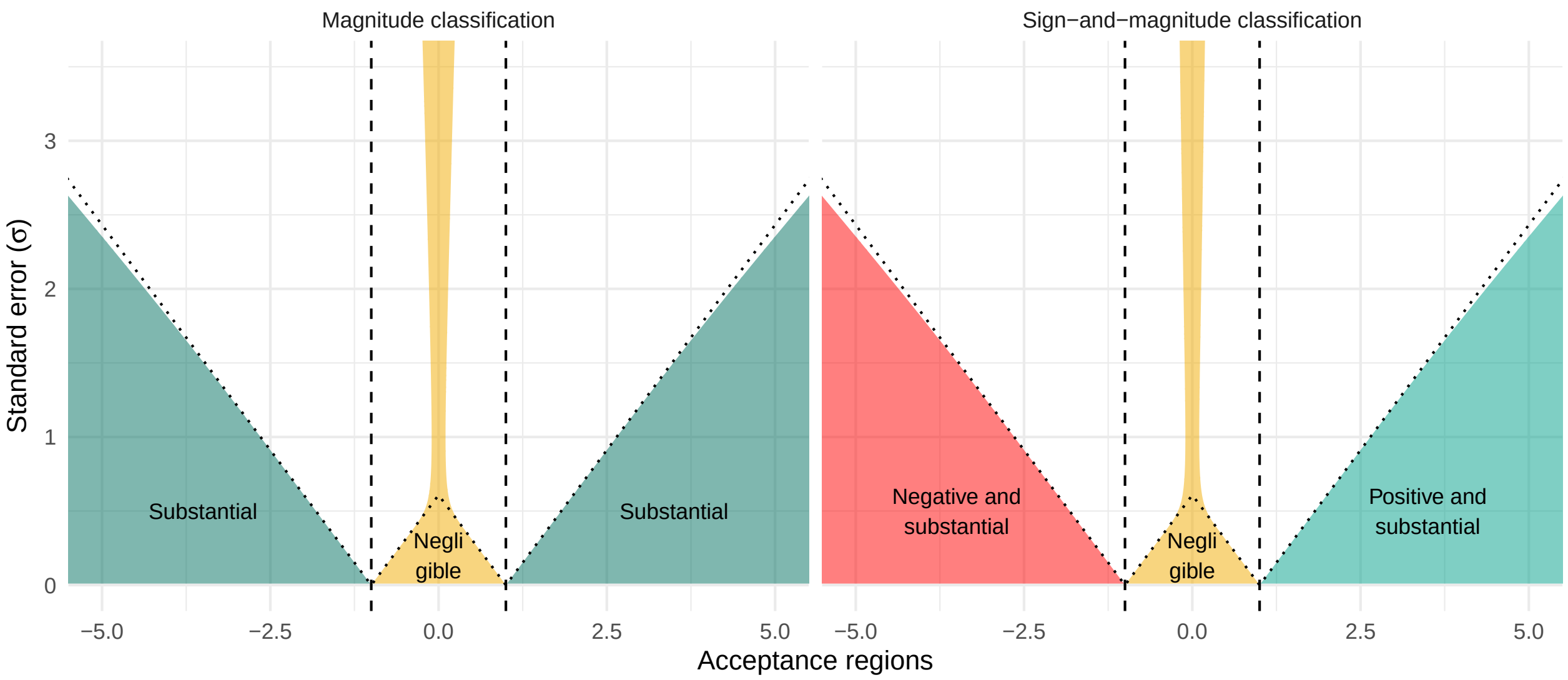

Figure 5: Acceptance regions for the magnitude classification test (left) and the sign-and-magnitude classification test (right) for $\delta = 1$ and $\alpha = .05$ as a function of the standard error (vertical axis). The inner boundary of the outer acceptance regions is identical across the two tests (and slightly further from the "negligible" region than $z_{1-\alpha}\sigma$). The "negligible" acceptance region is slightly narrower for the sign-and-magnitude test because, for imprecise tests, classification error for $\theta$ just above $\delta$ includes the lower acceptance region.

## 5.3 Sign-and-Magnitude Classification

In many theoretical and practical debates in social science, we are interested not only in an estimand's magnitude (small or not small) but also in its sign. For example, in a study comparing two approaches to mobilizing voters, it could be practically useful to determine whether intervention $A$ is substantially more effective than intervention $B$, intervention $B$ is substantially more effective than intervention $A$, or they are roughly equally effective.

For such cases, we consider a *sign-and-magnitude classification test* in which a bounded, "interior" class is flanked by one half-bounded class on each side. If the interior class is small and centered on zero, the classes are $C_0 = (-\delta, \delta)$ ("negligible"), $C_- = (-\infty, -\delta]$ ("negative and substantial"), and $C_+ = [\delta, \infty)$ ("positive and substantial"). If the interior class is centered at some other $c$, the estimand can be defined as $\theta - c$ so that the name fits.

Let $a$ refer to the inner limit of the acceptance region for the "positive and substantial" class; the symmetry of the problem implies that the inner limit of the acceptance region for the "negative and substantial" class is $-a$. For $\theta$ in the "negligible" class, a misclassification error occurs whenever $\hat{\theta} \geq a$ or $\hat{\theta} \leq -a$. As with the magnitude classification test, we therefore choose $a$ such that

$$P_{\theta=\delta}\left(\hat{\theta} \in (-\infty, -a] \cup [a, \infty)\right) = \alpha, \tag{3}$$

which caps at $\alpha$ the probability of misclassifying a borderline "negligible" $\theta$ as either "positive and substantial" or "negative and substantial".

As above, let $b$ refer to the upper limit of the acceptance region for the "negligible" class; given the symmetry of the problem, the lower limit is $-b$. We choose $b$ (given the choice of $a$) by pinning the worst-case misclassification rate for the two "substantial" classes at $\alpha$. The key distinction from the magnitude classification test is that we can misclassify $\theta = \delta$ (the lower boundary of the "positive and substantial" class) either as "negligible" (if $\hat{\theta}$ falls between $-b$ and $b$) or as "negative and substantial" (if $\hat{\theta}$ falls below $-a$). (We can make a corresponding statement for $\theta = -\delta$.) Thus we choose $b$ such that

$$P_{\theta=\delta}\left(\hat{\theta} \in (-b, b) \cup (-\infty, -a]\right) = \alpha, \tag{4}$$

which caps at $\alpha$ the probability of misclassifying a borderline "positive and substantial" $\theta$. Compared to the magnitude classification case, this extra misclassification mode requires a slightly smaller acceptance region for the "negligible" class.

The right panel of Figure 5 shows how the acceptance regions for the sign-and-magnitude test vary with test precision for fixed $\delta = 1$ and $\alpha = .05$. The outer acceptance regions are identical to those in the magnitude classification test (though here the upper and lower regions correspond to distinct conclusions). The acceptance region for the "negligible" class is slightly narrower for reasons explained above. As in the left panel, the dotted line shows the outline of the acceptance regions implied by using the 90% CI to classify $\theta$. Appendix Figure A2 illustrates how acceptance regions are determined for the sign-and-magnitude classification test.

Our sign-and-magnitude classification test is related to, but distinct from, the three-category tests proposed by Goeman et al. (2010) and Isager and Fitzgerald (2024), in which the thresholds $a$ and $b$ depend on $\sigma$ but not on $\delta$. While our approach invariably yields a misclassification error rate of $\alpha$ at both sides of the class boundaries, their fixed-threshold approach achieves this only at one value of $\delta$ for a given $\sigma$. Appendix Section E.1 shows via simulation the power advantage of our approach.

# 6 Combining the Results of Multiple Classification Tests

In many cases there may be more than one set of qualitative distinctions worth making. For example, we often would like to know both the sign of $\theta$ and whether it is substantial or negligible in magnitude. In such cases, a researcher can report side-by-side the error-controlled results of two or more classification tests for a given estimand $\theta$, simultaneously addressing two questions about that estimand. For example, nothing prevents us from reporting both that a sign test indicates $\theta > 0$ and that a magnitude test indicates $|\theta| < \delta$ (or a magnitude test is inconclusive), where each statement comes with its own error guarantee.

In this section we state conditions under which we can combine the conclusive results of two or more classification tests to make a more refined statement about $\theta$. We ask: If we have two or more conclusive tests, each placing $\theta$ in an interval, can we make an error-controlled statement that $\theta$ lies in the intersection of those intervals? The answer is that we can do so iff each bound in the intersection comes with its own error guarantee, though combining two tests is not generally the most powerful way to test whether $\theta$ is in that interval.

Consider first the straightforward case where the interval in the combined claim is identical to the interval of one of the component tests. For example, suppose that a size-$\alpha$ sign test asserts $\theta > 0$ and a size-$\alpha$ sign-and-magnitude test asserts $\theta \geq \delta > 0$. The intersection of the intervals is $\theta \geq \delta$. Because the sign-and-magnitude test "single-handedly" warrants that claim, its size-$\alpha$ error guarantee applies. All half-bounded intervals that result from combining multiple tests have this feature.

Next consider cases where the intersection of the results of two or more classification tests is a bounded interval between $l$ and $u$ and no single test places $\theta$ in that interval. Then the error guarantees of the underlying tests apply iff one test warrants the half-bounded claim that $\theta > l$ (or $\theta \geq l$) and another warrants the half-bounded claim that $\theta < u$ (or $\theta \leq u$). Thus the underlying error guarantees apply to the combined statement when each of $l$ and $u$ comes from a half-bounded claim: a sign test, or a "substantial and negative" or "substantial and positive" claim from a sign-and-magnitude test. They do not apply if either bound comes from a "negligible" claim or from any claim from a magnitude test.

It should be clear that separate error-controlled claims that $\theta > l$ and $\theta < u$ jointly imply the error-controlled claim that $\theta \in (l, u)$: one test limits the probability of mistakenly concluding that $\theta > l$, the other limits the probability of mistakenly concluding that $\theta < u$. It may also be clear from our critique of TOST above that the optimal classification test for determining whether $\theta \in (l, u)$ is based on a partition that includes that interval as a distinct class rather than (in the manner of TOST) combining two sign tests, one for each bound. As with TOST, the error bound is valid but the test will generally be conservative.

It may be less obvious that we cannot retain the error guarantee otherwise, i.e., if at least one of the bounds of the new interval is not from a half-bounded claim. Suppose we combined a sign test reporting $\theta > 0$ and a magnitude test (or a sign-and-magnitude test) reporting $|\theta| < \delta$ to conclude that $\theta \in (0, \delta)$. The problem is that the error control behind the test that supplied the upper bound applies to the claim that $|\theta| < \delta$, not to the claim that $\theta < \delta$. Similarly, suppose we combined a sign test reporting $\theta > 0$ with a magnitude test (*not* a sign-and-magnitude test) reporting $|\theta| \geq \delta$ to conclude that $\theta \geq \delta$. This intuitive move is not warranted, because the magnitude test result does not provide an error guarantee that $\theta \geq \delta$. (If it were warranted, there would be no reason to conduct a sign-and-magnitude test rather than both a sign test and a magnitude test, because the latter would allow "positive and substantial" and "negative and substantial" conclusions whenever the former did while also allowing "negligible" conclusions more often.)

In brief, we can confidently combine the conclusions from multiple classification tests when (i) the combined conclusion is exactly the same as the conclusion of one of the tests, or (ii) the combined conclusion locates $\theta$ in a bounded interval, where one bound is warranted by one test and the other is warranted by another. We apply these insights in interpreting the results of two classification tests in our application below.

# 7 Application

We illustrate the framework on the Broockman and Kalla (2025) field experiment, which studied the effect of exposing regular Fox News viewers to CNN. We set $\delta = 0.15$ standard deviations, a threshold below which average treatment effects would be considered substantively small (Cohen 1988). In general researchers should have a substantive rationale for defining negligible effects (Lakens 2017a; Rainey 2014) and other qualitative distinctions in their classification tests.

Figure 6 reports the results for all 32 pre-registered indices in four columns: a two-tailed test (the original testing approach),[18] the sign classification test, the sign-and-magnitude

[18] Our estimates and standard errors come from an R replication of Broockman and Kalla (2025)'s analysis pipeline, which uses elastic net regression to select covariates before OLS estimation. Because R's glmnet

classification test, and the combination of the two classification tests.[19]

The two-tailed test (first column) yields 16 significant and 16 non-significant results. The sign test (second column) classifies 18 estimands by direction (17 as positive and one as negative); it signs every significant result and two more, because it uses the threshold $z_{1-\alpha} \approx 1.645$ in place of $z_{1-\alpha/2} \approx 1.96$ in the two-tailed test. The sign-and-magnitude test (third column) classifies two indices as "positive and substantial" and 12 as "negligible", and returns "inconclusive" for the remaining eighteen.

The final column reports the sharpest error-controlled claim that the sign and sign-and-magnitude results jointly warrant. (See Section 6.) Where the sign-and-magnitude test returns a substantial claim, that claim implies the sign claim ($\theta \geq \delta \implies \theta > 0$), so the combined claim is the one from the sign-and-magnitude test; this occurs for two indices. Where only one test is conclusive, the combined claim is that test's conclusion. One index (Dem. Political Preferences) is conclusive under both tests without a mergeable bound: the sign test warrants $\theta > 0$ and the sign-and-magnitude test warrants $|\theta| < \delta$, but the combining rules do not permit the bounded claim $\theta \in (0, \delta)$. We report the negligible result for that index. Combining the two tests produces an error-controlled classification for 29 of the 32 indices ($29/32 \approx .91$), against the 16 ($16/32 = .50$) for which the two-tailed test permits an error-controlled claim that $\theta \neq 0$.

Together, the classification tests add error-controlled signs for originally significant and insignificant effects and "negligible" designations to some insignificant effects, and identify effects that are not just significant but substantial in magnitude. The application thus illustrates the richer qualitative claims that classification testing can provide compared to the standard two-tailed test.

and Stata's elasticregress select different covariates, a few borderline outcomes differ from those reported in Broockman and Kalla (2025).

[19]We apply no correction for multiple testing to any column, so each classification controls the error rate at $\alpha = 0.05$ for its own estimand. The original study corrects for multiple testing within survey wave. In Appendix Section B and Section C, we develop classification p-values and show how they can be used in standard corrections for multiple testing when a summary conclusion depends on many estimands at once.

|  |  | Two−tailed test | Sign classification test | Sign−and−magnitude classification test | Combined claim |
|---|---|---|---|---|---|
| Midline | Substitute News Source Use | Significant | Negative | Inconclusive | Negative |
| Midline | Extremity Covered Issues | Not significant | Inconclusive | Inconclusive | Inconclusive |
| Midline | Fox−Covered Trump Positions | Not significant | Inconclusive | Negligible | Negligible |
| Midline | General Media Attitudes | Not significant | Inconclusive | Negligible | Negligible |
| Midline | Second Order Beliefs | Not significant | Inconclusive | Negligible | Negligible |
| Midline | Reduced Ethnic Antagonism | Not significant | Inconclusive | Negligible | Negligible |
| Midline | Favorable CNN Attitudes | Not significant | Inconclusive | Negligible | Negligible |
| Midline | Support for Dem. Norms | Not significant | Inconclusive | Negligible | Negligible |
| Midline | Reduced Racial Prejudice | Not significant | Inconclusive | Negligible | Negligible |
| Midline | Affect towards Dem. Voters | Not significant | Inconclusive | Negligible | Negligible |
| Midline | CNN−Covered Biden Positions | Not significant | Inconclusive | Inconclusive | Inconclusive |
| Midline | Biden Evaluation | Not significant | Positive | Inconclusive | Positive |
| Midline | Dem. Political Preferences | Significant | Positive | Negligible | Negligible |
| Midline | Reduced Fox Viewership | Not significant | Positive | Inconclusive | Positive |
| Midline | Liberal Pref. Non−Covered Issues | Not significant | Inconclusive | Inconclusive | Inconclusive |
| Midline | Affect towards Rep. Voters | Significant | Positive | Inconclusive | Positive |
| Midline | COVID Attitudes | Significant | Positive | Inconclusive | Positive |
| Midline | Unfavorable Fox Attitudes | Significant | Positive | Inconclusive | Positive |
| Midline | CNN Covered Perceptions (Non−COVID) | Significant | Positive | Inconclusive | Positive |
| Midline | Liberal Pref. Covered Issues | Significant | Positive | Inconclusive | Positive |
| Midline | Fox−Covered Biden Positions | Significant | Positive | Inconclusive | Positive |
| Midline | Reduced Trump Evaluation | Significant | Positive | Inconclusive | Positive |
| Midline | CNN−Covered Trump Positions | Significant | Positive | Inconclusive | Positive |
| Midline | CNN vs. Fox Issue Importance | Significant | Positive | Inconclusive | Positive |
| Midline | Fox Covered Perceptions (Non−COVID) | Significant | Positive | Inconclusive | Positive |
| Quiz | Fox Trust (Quiz) | Not significant | Inconclusive | Negligible | Negligible |
| Quiz | General Media Attitudes (Quiz) | Not significant | Inconclusive | Negligible | Negligible |
| Quiz | Short−Term Emotion (Quiz) | Not significant | Inconclusive | Negligible | Negligible |
| Quiz | CNN Trust (Quiz) | Significant | Positive | Inconclusive | Positive |
| Quiz | Attitudes Towards Events (Quiz) | Significant | Positive | Inconclusive | Positive |
| Quiz | Partisan Valence (Quiz) | Significant | Positive | Pos. and substantial | Pos. and substantial |
| Quiz | Current Event Perceptions (Quiz) | Significant | Positive | Pos. and substantial | Pos. and substantial |

Figure 6: Classification of the 32 pre-registered indices from Broockman and Kalla (2025) under two-tailed test, the sign test, the sign-and-magnitude test, and the combination of the two classification tests. Each classification controls the error rate at $\alpha = 0.05$ per estimand ($\delta = 0.15$ SD). The sign-and-magnitude test distinguishes "positive and substantial" ($\theta \geq \delta$), "negligible" ($|\theta| < \delta$), and "negative and substantial" ($\theta \neq -\delta$). The combined column reports the sharpest error-controlled claim that the sign and sign-and-magnitude results jointly warrant. "Inconclusive" means the evidence is insufficient to classify $\theta$. For the one index whose sign and sign-and-magnitude claims are separately conclusive but cannot be merged into a bounded interval, the combined column reports the sign-and-magnitude result.

# 8 How to Implement a Classification Test

The first step in classification testing is designing the test. (See box.) Given an estimand $\theta$, estimator $\widehat{\theta}$, and estimator for the standard error $\widehat{\sigma}$, the key decision is how to partition the parameter space of $\theta$ into qualitatively distinct classes. In cases where attention focuses on the estimand's sign, a sign classification test is an obvious choice, and can be combined with other tests as discussed above. In other cases the researcher must apply more detailed substantive knowledge and justify the partition. Choosing quantitative cutoffs to demarcate qualitative distinctions is a familiar problem from the design of equivalence tests, where researchers must decide how large a "negligible" effect can be (Lakens 2017a; Rainey 2014), and is thus not unique to classification tests.

**Implementing a classification test**

1. Design stage (pre-registration)
    a. Define the estimand $\theta$, estimator $\widehat{\theta}$, and estimator for the standard error $\widehat{\sigma}$
    b. Choose an error level $\alpha$
    c. Partition $\theta$ into two or more classes
2. Implementation stage
    a. Obtain estimates $\widehat{\theta}$ and $\widehat{\sigma}$
    b. Classify $\theta$ if it falls in an acceptance region (constructed "from scratch" or using our software), or declare inconclusive

A classification test is more likely to be persuasive if it is pre-registered. Otherwise, the research audience may suspect that the partition of $\theta$ (or the use of classification testing instead of conventional hypothesis tests) was chosen for instrumental rather than scientific reasons.[20] If the researcher plans to classify multiple estimands and arrive at a summary conclusion with error control, she should state this mapping along with a method of using

[20] Journal editors and other researchers find one-tailed tests more acceptable when they are pre-registered (Hales 2024; Lakens 2017b).

classification p-values to adjust for multiplicity (Appendix Section B and Section C).

Once the test is designed and estimates are obtained, implementation is very simple for the three standard tests we have described: given a partition, an $\alpha$, an estimate, and a standard error, our software returns the classification along with a p-value (see Appendix Section B). Tests with finer partitions can be designed, but (i) the advantage over reporting raw estimates becomes smaller and (ii) determining acceptance region boundaries may raise questions not resolved in this paper.[21]

# 9 Discussion and Conclusion

Classification tests improve on current practice by allowing researchers to either adjudicate between hypotheses or test one in a more demanding way. The sign classification test delivers directional claims at no cost relative to a one-tailed test; the magnitude classification test allows improved detection of negligible effects (compared to standard equivalence tests) while also permitting non-equivalence conclusions (like the ubiquitous two-tailed test); and the sign-and-magnitude classification test distinguishes negligible effects from substantial effects in each direction. Our re-analysis of Broockman and Kalla (2025) shows how switching from a standard two-tailed significance test to classification testing can produce richer error-controlled conclusions.

We have argued that classification tests are superior to current practice in part because they allow researchers to draw error-controlled conclusions in more circumstances, including when surprising results are observed. Some readers may, however, feel that existing testing procedures are already too permissive, such that we should be wary of making it still easier to obtain affirmative conclusions. We offer two responses to this concern.

First, if the community seeks stronger error guarantees, the correct response is to lower the nominal level $\alpha$ and take steps to ensure that the actual error rate is capped at the nominal level (e.g., through pre-registration and better peer review). Instead of deliberately choosing less powerful procedures, we should ensure that error rates are controlled and use the most powerful procedure available.

[21] Specifically, with more than three compact classes, the test design is no longer uniquely determined by the error rate at the class boundaries; one must choose among several designs, e.g., by maximizing expected power under a prior distribution for $\theta$.

Second, we suggest that concerns about "permissiveness" are at least partly misplaced, because they are responses to incentive problems that classification testing helps address. Current practice allows affirmative results only in the direction of the pre-selected research hypothesis, so researchers seek affirmative results and face pressure to obtain results consistent with their chosen hypothesis (i.e., to avoid a "null result"). This pressure invites inappropriate research behavior, which pushes the community to embrace constraints like pre-registration and multiple testing corrections that make it harder for researchers to (unjustifiably) claim an affirmative result. In this atmosphere of suspicion, a testing method that allows for more affirmative conclusions seems like a step backward. But classification testing may reduce the incentive to behave inappropriately (and thus the need for more restrictive testing procedures), because it expands the set of error-controlled conclusions that are ex ante possible and discourages the community from seeing research as a test of the researcher's prognostic ability. Researchers using classification tests will still hope to avoid inconclusive results, and they may prefer certain conclusions for ideological or theoretical reasons. But to the extent that classification testing makes researchers feel less invested in obtaining specific findings, it will address one of the problems that motivate concerns about "permissiveness".

Another possible concern is that something valuable would be lost if researchers stopped pre-registering a research hypothesis (as is currently common) and instead began pre-registering only a set of rival possibilities (via classification testing). Again we offer two responses.

First, nothing prevents a researcher who uses classification tests from stating a research hypothesis as they would currently. Classification testing decouples test design from the choice of hypothesis, but it does not prevent a researcher from advancing (and pre-registering) a hypothesis if it is valuable to do so.

Second, we suggest that pre-registering hypotheses is less valuable (and more harmful) than many researchers believe. We suspect that researchers pre-register a research hypothesis not just because current hypothesis-testing practice seems to require it but also to satisfy an expectation that they can predict the experiment's outcome. But an experiment where no other outcome is plausible is not worth conducting. We learn most from studying situations where systematic forces are plausibly at work, yet the outcome remains uncertain despite our theories and prior empirical research (Chamberlin 1890; Clarke 2007; Gigerenzer et al. 2004; Platt 1964). Moreover, as noted above, when researchers have offered a prediction, they face pressure to obtain findings consistent with that prediction, which creates an incentive to distort the outcome. In this sense it should be viewed as a feature of a good design that the researcher can clearly articulate what we will learn from the (ex ante uncertain) outcome of a study but refrains from predicting that outcome.

# Online Appendix

## Appendix Contents

# A Determining the Acceptance Regions

In this section, we show how to determine the acceptance regions for the magnitude and the sign-and-magnitude tests.

## A.1 Determining Acceptance Regions in the Magnitude Classification Test

Figure A1 shows how acceptance regions are determined in a magnitude classification test for a case where the estimator is precise (top) and imprecise (bottom) compared to the width of the "negligible" region. When the estimator is precise, all acceptance region boundaries are located roughly $z_{1-\alpha}\sigma$ from class boundaries. Thus the "negligible" acceptance region is roughly $[-\delta + z_{1-\alpha}\sigma, \delta - z_{1-\alpha}\sigma]$ (as in TOST). This is because the left tail of the sampling distribution when $\theta = \delta$ lies almost entirely within the "negligible" class's acceptance region, so that we essentially have a sign classification problem at each boundary point $\delta$ and $-\delta$. When the estimator is less precise (lower figure), the error budget for $\theta = \delta$ (considered as part of the "negligible" class) is divided across both parts of the "substantial" class's acceptance region. As a result, these outer acceptance regions must be slightly further from the class boundaries than $z_{1-\alpha}\sigma$. The boundaries of the "negligible" class's acceptance region, by contrast, are *closer* to the class boundaries than $z_{1-\alpha}\sigma$. This is because the error budget for $\theta$ just above $\delta$ is not spent entirely in the tails of the sampling distribution but closer to the mode. For an estimator this imprecise, there would be no rejection region via TOST, highlighting the conservatism of that equivalence test.

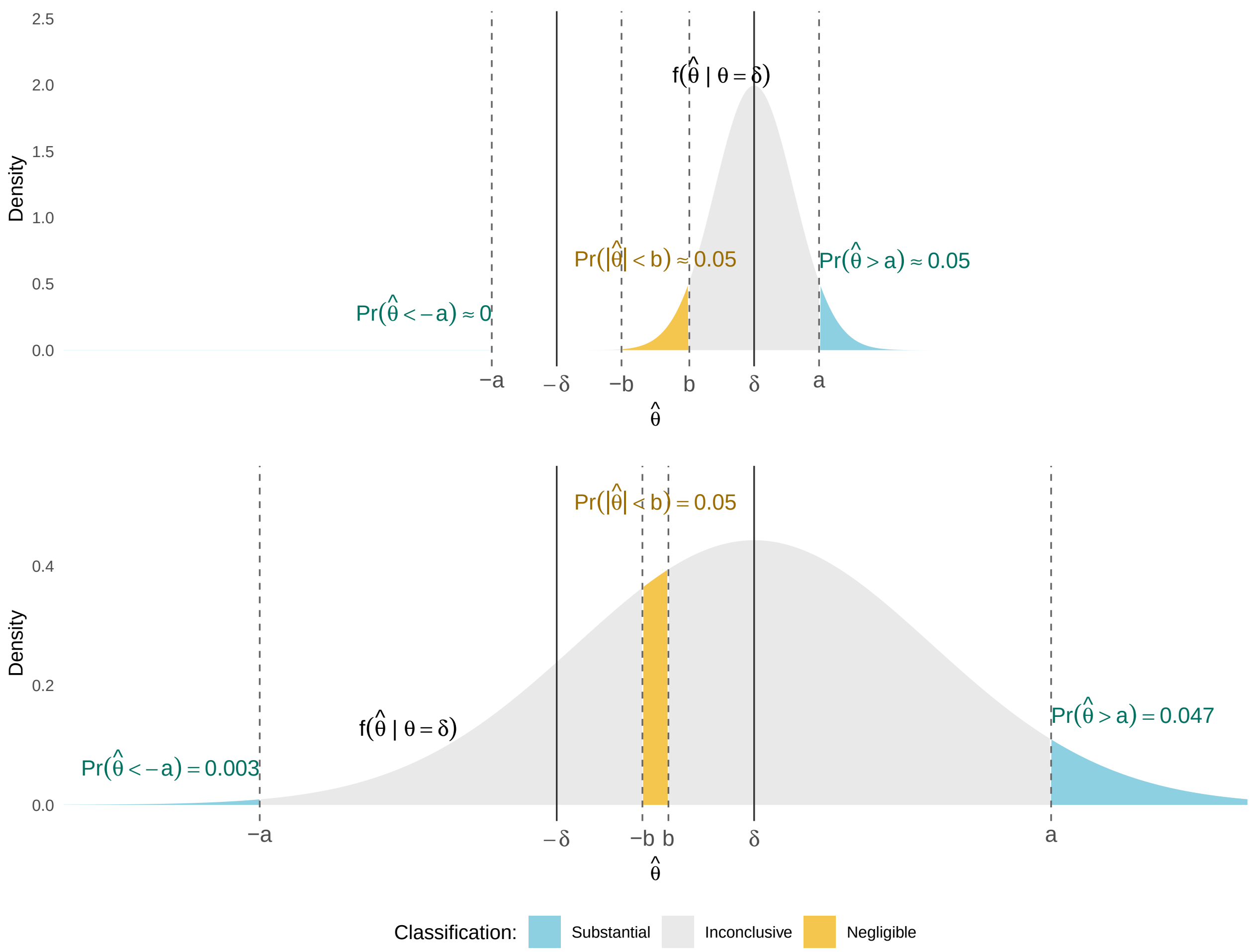

Figure A1: Classifications at $\theta = \delta$ for the magnitude classification test: precise estimation (top) and imprecise estimation (bottom). The inner boundaries of the acceptance regions for the "substantial" class ($a$ and $-a$, by symmetry) are set so that, when $\theta$ is at the upper boundary of the "negligible" class (i.e., $\delta$), the probability of an incorrect classification ($\widehat{\theta} \geq a$ or $\widehat{\theta} \leq -a$) is exactly $\alpha$ (here, 0.05). The boundaries of the acceptance region for the "negligible" class ($b$ and $-b$, by symmetry) are set so that, when $\theta$ is at the inner boundary of the "substantial" class (i.e., the limit as $\theta$ goes to $\delta$ from above), the probability of an incorrect classification ($\widehat{\theta} \in (-b, b)$) is exactly $\alpha$. An "inconclusive" result occurs when $\widehat{\theta}$ falls outside of those acceptance regions. In the precise estimation case, $a \approx \delta + 1.645\sigma$ and $b \approx \delta - 1.645\sigma$; the critical $z$s are about the same as in the sign classification test. In the imprecise case, $a \approx \delta + 1.67\sigma$ (where the critical $z$ is larger because of the risk of misclassification in the other direction) and $b \approx \delta - 0.48\sigma$ (where the critical $z$ is smaller because the acceptance region is no longer in the tail of the sampling distribution).

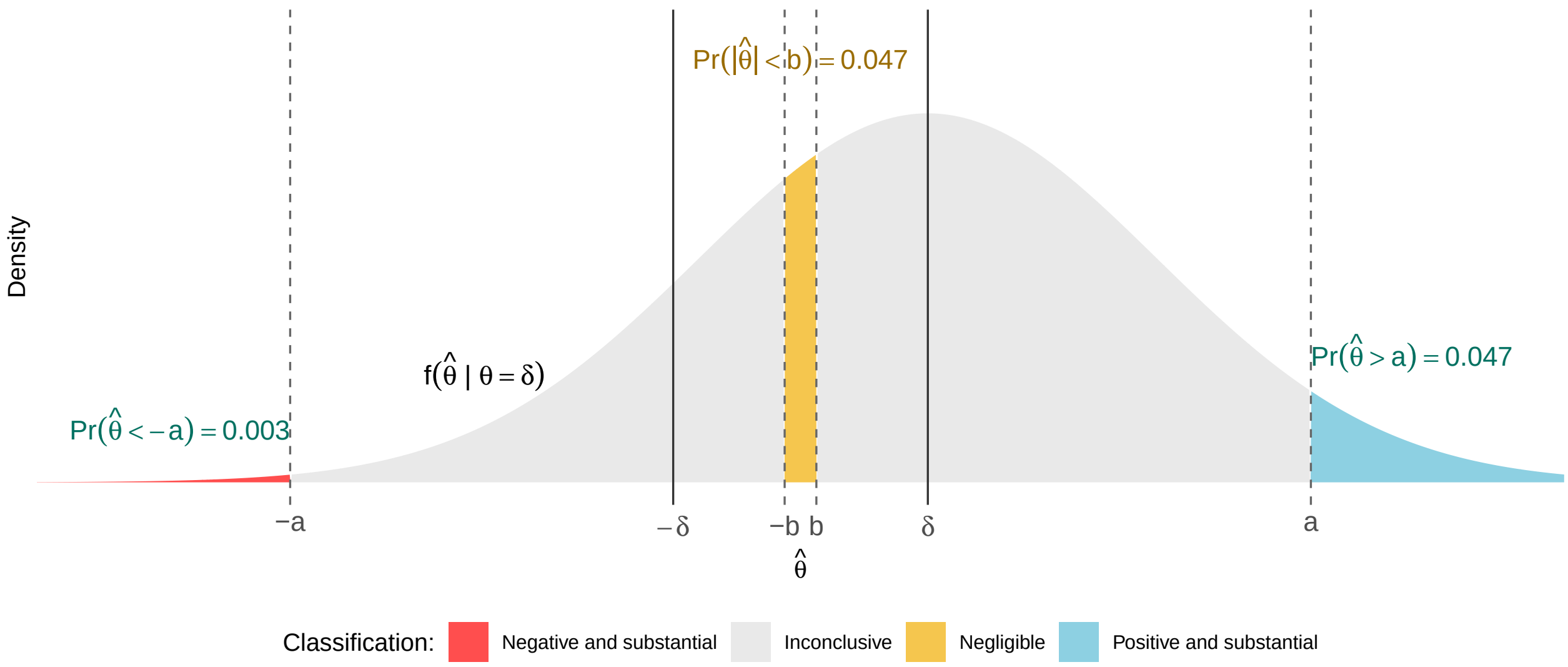

Figure A2: Classifications at the boundary $\theta = \delta$ for the sign-and-magnitude classification test (imprecise estimation). The inner boundaries of the acceptance regions for the "positive and substantial" class ($a$) and the "negative and substantial" class ($-a$, by symmetry) are set so that, when $\theta$ is at the upper boundary of the "negligible" class (i.e., $\delta$), the probability of an incorrect classification ($\hat{\theta} \geq a$ or $\hat{\theta} \leq -a$) is exactly $\alpha$ (here, 0.05). The boundaries of the acceptance region for the "negligible" class ($b$ and $-b$, by symmetry) are set so that, when $\theta$ is at the inner boundary of the "positive and substantial" class (i.e., the limit as $\theta$ goes to $\delta$ from above), the probability of an incorrect classification ($\hat{\theta} \in (-b, b) \cup (-\infty, -a]$) is exactly $\alpha$. An "inconclusive" result occurs when $\hat{\theta}$ falls outside of those acceptance regions. Compared to the boundaries in the magnitude classification test, $a$ is the same but $b$ is slightly smaller (here, $b \approx \delta - 0.49\sigma$) because a very low $\hat{\theta}$ now produces a misclassification when $\theta \geq \delta$.

### A.2 Determining Acceptance Regions in the Sign-and-Magnitude Classification Test

Figure A2 illustrates the process of determining acceptance regions for the sign-and-magnitude classification test. As with the magnitude classification test, if estimation is precise (as in the top panel of Figure A1), all acceptance region boundaries are approximately $z_{1-\alpha}\sigma$ from a class boundary. We show instead an imprecise case far from this benchmark. The acceptance regions of the outer classes are chosen to pin the error rate at $\theta = \delta$ (the upper boundary of the "negligible" class) at $\alpha$, exactly as in Figure A1. The acceptance region of the "negligible" class is chosen so that the error where $\theta$ is just above $\delta$ (and thus barely part of the "positive and substantial" class) is pinned at $\alpha$. This error combines the mass in both of the other acceptance regions, not just the "negligible" one, which requires the "negligible" acceptance region to be slightly narrower than in the magnitude classification test.

## B Classification P-Values

Apart from the classification, researchers may also want a measure of how decisively an estimand can be assigned to a class based on a particular estimate. We define such a measure, which we call a "classification p-value". The classification p-value $\underline{p}$ is the smallest target error rate at which a quantity receives a definitive classification. A quantity with $\underline{p} = 0.02$ would be classified definitively at any target rate $\alpha \geq 0.02$. A quantity with $\underline{p} = 0.08$ would be Inconclusive at $\alpha = 0.05$ but classified definitively at $\alpha = 0.08$. Smaller values indicate stronger evidence for the assigned classification. We recommend that researchers report $\underline{p}$ alongside the classification, just as they report a p-value alongside a significance decision.

One reason to calculate classification p-values is to correct for multiple comparisons when a single decision (e.g., declaring that a claim has been supported) rests on several classifi-cation tests, each involving a distinct estimand. The researcher can then use the p-values from each classification test as inputs to standard procedures that control the false discovery rate (e.g., Benjamini and Hochberg (1995)) or the family-wise error rate (e.g., Holm (1979)), which we describe in Section C. Those procedures require a single property of the classification p-value, established below: at any target error rate $t$, a definitive but wrong classification has probability at most $t$.

Formally, let $C(\hat{\theta}, \sigma, \delta;\, \alpha)$ denote the classification produced at target error rate $\alpha$ (where $\delta$ summarizes the partitioning). Then

$$\underline{p} = \inf\{\alpha \in (0,1) : C(\hat{\theta}, \sigma, \delta;\, \alpha) \neq \text{Inconclusive}\}.$$

Because the acceptance regions are larger at a higher $\alpha$, the procedure classifies definitively if and only if $\underline{p} \leq \alpha$, so reporting $\underline{p}$ is equivalent to reporting the classification at every $\alpha$.

For the sign classification test, the classification p-value equals the conventional one-sided p-value in the observed direction: $\underline{p}_{\text{sign}} = \Phi(-|\hat{\theta}|/\sigma)$. (This follows because the sign classification test classifies whenever $|\hat{\theta}|/\sigma > z_{1-\alpha}$, so the smallest such $\alpha$ satisfies $\Phi^{-1}(1-\alpha) = |\hat{\theta}|/\sigma$.) For the magnitude and sign-and-magnitude classification tests, the smallest $\alpha$ at which $\hat{\theta}$ lies in an acceptance region has no closed form and is computed numerically. Each test's classification p-value is bounded above (by $1/2$ for the sign and magnitude tests; see Property 2 below), so at any target rate at or above that bound every estimand receives a definitive classification.

## B.1 Properties of Classification P-Values

We establish two properties of classification p-values: Property 1, on which the multiplicity corrections of Section C rely, and Property 2, a bound on how large a classification p-value can be.

**Property 1.** For each classification test and every target error rate $t \in (0, 1/2]$, a definitive but wrong classification, one made at rate $t$ (and thus with $\underline{p} \leq t$), has probability at most $t$. This probability reaches $t$ as $\theta$ approaches a class boundary:

$$\sup_{\theta} P_{\theta}(\underline{p} \leq t \text{ and the classification is wrong}) = t.$$

Just as $P_{\theta}(p \leq t) \leq t$ under a true null makes the rule that rejects when $p \leq t$ control the false-rejection rate at $t$, the bound above makes the rule that classifies when $\underline{p} \leq t$ control the misclassification rate at $t$.

*Proof.* In each test's calibration, we set its worst-case misclassification probability to the target error rate, and that worst case occurs at a class boundary: at $\theta = 0$ for the sign test (Proposition 3), and as $\theta \to \pm\delta$ for the magnitude and sign-and-magnitude tests (Propositions 4 and 2). Evaluated at target error rate $t$, each result gives a misclassification probability equal to $t$ at the boundary and strictly smaller for $\theta$ in non-boundary region(s) of a class, so the supremum over $\theta$ is $t$. This misclassification probability at rate $t$ is $P_{\theta}(\underline{p} \leq t$ and the classification is wrong), because $\{\underline{p} \leq t\}$ is the event that $\hat{\theta}$ is classified definitively at rate $t$. □

**Property 2 (Bound on the classification p-value).** Each test's classification p-value has a finite supremum over estimates: at most $1/2$ for the sign and magnitude tests, and slightly above $1/2$ for the sign-and-magnitude test. Writing $z = \hat{\theta}/\sigma$ and $r = \delta/\sigma$:

(i) *Sign test.* $\underline{p}_{\text{sign}} = \Phi(-|z|) \leq 1/2$, with equality iff $z = 0$.

(ii) *Magnitude test.* $\sup_z \underline{p}_{\text{mag}} = 1/2$ for every $r$, attained at $|z| = u^*(r)$, the unique solution of $\Phi(u - r) - \Phi(-u - r) = 1/2$.

(iii) *Sign-and-magnitude test.* $\sup_z \underline{p}_{\text{sam}} = \frac{1}{2} + \frac{1}{2}\Phi(-2r - z_d)$, where $z_d > 0$ solves $\Phi(z_d) = \frac{1}{2} + \frac{1}{2}\Phi(-2r - z_d)$. The bound exceeds $1/2$ and, as $r \to \infty$, decreases to $1/2$.

*Proof.*

(i) $\Phi$ is increasing, so $\Phi(-|z|)$ is largest when $|z|$ is smallest. At $z = 0$, $\Phi(0) = 1/2$. For $z \neq 0$, $\Phi(-|z|) < 1/2$.

(ii) The magnitude test classifies $\widehat{\theta}$ as "substantial" when $|\widehat{\theta}| \geq a$ and as "negligible" when $|\widehat{\theta}| < b$, with $a$ and $b$ calibrated at level $\alpha$ by Equation 1 and Equation 2. In units of $z$ the substantial threshold is $a/\sigma$ and the negligible threshold is $b/\sigma$. As $\alpha$ increases, $a/\sigma$ decreases and $b/\sigma$ increases, so both acceptance regions are larger and a fixed $z$ receives a definitive classification at the smallest $\alpha$ for which $a/\sigma = |z|$ or $b/\sigma = |z|$. The supremum of $\underline{p}_{\text{mag}}$ over $z$ is therefore the level at which the inconclusive interval $(b/\sigma,\, a/\sigma)$ becomes empty (that is, at which $a = b$). Writing Equation 1 and Equation 2 in units of $z$ with the common value $u = a/\sigma = b/\sigma$ gives

$$1 - \Phi(u - r) + \Phi(-u - r) = \alpha \qquad \text{and} \qquad \Phi(u - r) - \Phi(-u - r) = \alpha.$$

Adding the two equations gives $1 = 2\alpha$, so the inconclusive interval becomes empty at $\alpha = 1/2$. Substituting $\alpha = 1/2$ into the second equation gives $\Phi(u-r)-\Phi(-u-r) = 1/2$, whose left-hand side increases strictly from $0$ at $u = 0$ to $1$ as $u \to \infty$. By the intermediate value theorem the solution $u^*(r)$ is unique. For every $\alpha < 1/2$ the first condition gives $a/\sigma > u^*(r)$ and the second gives $b/\sigma < u^*(r)$, so the interval $(b/\sigma,\, a/\sigma)$ is non-empty and contains $u^*(r)$. At $\alpha = 1/2$ its endpoints coincide at $u^*(r)$. Hence $\sup_z \underline{p}_{\text{mag}} = 1/2$, attained at $|z| = u^*(r)$.

(iii) The sign-and-magnitude test classifies $\widehat{\theta}$ as "positive and substantial" or "negative and substantial" when $|\widehat{\theta}| \geq a = \delta + z_{\text{dir}}\sigma$ and as "negligible" when $|\widehat{\theta}| < b = \delta - z_{\text{eq}}\sigma$, with $(z_{\text{dir}}, z_{\text{eq}})$ calibrated at level $\alpha$ by Equation 3 and Equation 4. As in part (ii), a larger $\alpha$ yields larger acceptance regions (raising $\alpha$ relaxes Equation 3 and Equation 4, so the substantial thresholds decrease and the negligible interval is wider), so $\underline{p}_{\text{sam}}$ reaches its supremum over $z$ at the level $\alpha^*$ where the inconclusive interval becomes empty. By Step 3 of Proposition 5, the interval becomes empty when $z_{\text{eq}} = -z_{\text{dir}}$: at that level the three acceptance regions become adjacent and partition the line, so $P_+ + P_- + P_0 = 1$. Write $z_d := z_{\text{dir}}(\alpha^*) > 0$, so $z_{\text{eq}}(\alpha^*) = -z_d$. Constraint A (Equation 3) is $[1 - \Phi(z_d)] + \Phi(-2r - z_d) = \alpha^*$. At $z_{\text{eq}} = -z_d$ the misclassification probability for "negligible" is $P_0(-z_d) = \Phi(z_d) - \Phi(-2r - z_d)$, so Constraint B, $P_0(z_{\text{eq}}) + \Phi(-2r - z_{\text{dir}}) = \alpha^*$ (Equation 4), reduces to $\Phi(z_d) = \alpha^*$. Equating the two expressions for $\alpha^*$,

$$1 - \Phi(z_d) + \Phi(-2r - z_d) = \Phi(z_d) \quad \iff \quad \Phi(z_d) = \tfrac{1}{2} + \tfrac{1}{2}\Phi(-2r - z_d),$$

which has a unique solution $z_d > 0$ because the left-hand side increases strictly from $1/2$ at $z_d = 0$ to 1 while the right-hand side decreases from $\frac{1}{2} + \frac{1}{2}\Phi(-2r) < 1$. The bound is $\alpha^* = \Phi(z_d) = \frac{1}{2} + \frac{1}{2}\Phi(-2r - z_d)$. Because $\Phi(-2r - z_d) > 0$, the bound exceeds $1/2$. As $r \to \infty$, $\Phi(-2r - z_d) \to 0$, so $z_d \to 0$ and $\alpha^* \to 1/2$. $\square$

# C Multiplicity Corrections

The choice among correction procedures is independent of which classification test is used. In this section, we briefly discuss how to use common methods for correcting for multiple classification tests. The relevant scenario is that the researcher classifies $K$ different estimands and wants a summary conclusion (for example, "the treatment moved the average for at least one of the $K$ outcomes"). In such cases, the probability of making an error in the summary conclusion may be well above the probability of error in each component test. Classification p-values let researchers apply standard procedures for controlling the false discovery rate (FDR) or family-wise error rate (FWER) to this multiplicity problem.

FWER bounds the probability that any classification in the set is wrong, and should be used when even a single false classification would be costly. FDR controls the expected proportion of wrong conclusions among the conclusive ones, and should be used when researchers find it acceptable to classify most but not all quantities correctly in expectation. Because FWER bounds the probability of any mistake while FDR bounds only the expected proportion, FWER procedures typically classify fewer quantities as conclusive.

## C.1 Controlling the False Discovery Rate (FDR)

For a set of $K$ classification p-values, standard FDR procedures require the joint bound $P(\underline{p} \leq t \text{ and the classification is wrong}) \leq t$, which Property 1 (Appendix Section B) provides, with equality at the boundary. Because the false-discovery proportion counts only wrong classifications (instead of the falsely rejected nulls in significance testing) in its numerator, substituting the joint bound for the uniformity of a true null lets an FDR procedure control the FDR. Controlling FDR at level $q$ then guarantees that, on average, no more than a fraction $q$ of the conclusive classifications are wrong.

The Benjamini-Hochberg (BH) procedure (Benjamini and Hochberg 1995) orders the $K$ classification p-values as $\underline{p}_{(1)} \leq \underline{p}_{(2)} \leq \cdots \leq \underline{p}_{(K)}$, finds the largest $k$ such that $\underline{p}_{(k)} \leq qk/K$, and classifies all quantities with $\underline{p} \leq \underline{p}_{(k)}$ according to their per-comparison classification. Quantities above the threshold cannot be conclusively classified. Under independence, BH controls the FDR at level $q$.

When the $K$ estimators are correlated,[22] the classification p-values are dependent, and this dependence could in principle undermine FDR control. We assess it by simulation in Section C.3. In those simulations BH's realized FDR stays at or below the target even under a dependence structure designed to be adversarial for the procedure.

We can improve the power of BH using the two-stage procedure of Benjamini et al. (2006) (hereafter BKY). In stage 1, BH is applied at level $q/(1+q)$. The number of quantities left inconclusive in stage 1, $\hat{m}_0 = K - R_1$ (where $R_1$ is the number of quantities classified definitively), estimates how many quantities are truly at the classification boundary. In

[22] as is probably the case in the Broockman and Kalla (2025) application where outcome indices share a common treatment-control comparison

stage 2, BH is applied at the inflated level $[q/(1+q)] \cdot K/\hat{m}_0$. Because $\hat{m}_0 \leq K$, the stage-2 threshold is at least as large as the stage-1 threshold, admitting more conclusive classifications. Under independence, the two-stage procedure controls the FDR at $q$. The BKY procedure is the default multiplicity adjustment in several recent experimental studies in political science (Broockman and Kalla 2025; Westwood et al. 2022), following the Stata implementation by Anderson (2008).

## C.2 FWER Control

We discuss two procedures for controlling FWER. The Bonferroni procedure, which applies a flat threshold of $\alpha/K$ to each p-value, controls FWER by the same property: the probability that some classification is both wrong and conclusive is at most $\sum_{k=1}^{K} P(\underline{p}_k \leq \alpha/K \text{ and } k \text{ wrong}) \leq K \cdot (\alpha/K) = \alpha$, by the union bound and Property 1.

The Holm procedure (Holm 1979) is a step-down procedure on ordered p-values. Applying it to classification tests, we first order the $K$ classification p-values from smallest to largest. We then classify quantity $(i)$ conclusively if $\underline{p}_{(j)} \leq \alpha/(K-j+1)$ for all $j \leq i$. The procedure steps down the ordered list and stops at the first p-value that exceeds its threshold. All quantities at or above that point cannot be classified conclusively (i.e., they are "Inconclusive"). Holm is less conservative than Bonferroni's flat $\alpha/K$ threshold and controls FWER under arbitrary dependence.[23]

## C.3 Simulation Evidence

Figure A3 reports the realized FDR and FWER of the sign and sign-and-magnitude tests under different correction procedures. We draw $K = 30$ estimates $\hat{\theta}_k \sim N(\theta_k, 1)$ with correlation $\rho$, use each procedure to select which estimands are reported, and classify every selected estimand at the fixed per-comparison level $\alpha = 0.05$. A misclassification occurs when an asserted class excludes the true $\theta_k$. Each panel shows the realized rate of misclassification across 20,000 replications, with error bars indicating the realized rate $\pm 2$ Monte Carlo standard errors. We use the worst-case configuration for each test: every $\theta_k$ at the class boundary $\theta = 0$ for the sign test, and every $\theta_k$ just below $\delta$ for the sign-and-magnitude test (the boundary between the negligible and the positive-and-substantial class).

[23]This depends primarily on Property 1 (Appendix Section B).

Both the sign and sign-and-magnitude tests control the FDR and the FWER at the target across the dependence range. The realized rate is largest at independence, where it approaches the target rate, and decreases as $\rho$ rises.

To test the procedures' robustness to other forms of dependence, we repeat the exercise with four more data-generating processes (DGPs). All DGPs have the same $N(\theta_k, 1)$ marginals but different joint dependence. The four structures are: (1) positive equicorrelation (correlation $+\rho$), (2) signed blocks (correlation $+\rho$ within a block and $-\rho$ across blocks), (3) a t-copula (df 3) with tail dependence, and (4) an adversarial mixture. For the adversarial setting, we draw $K$ estimates from a mixture in which all $K$ estimates share one shock with a probability equal to the dependence strength. This structure is adversarial because the wrong classifications cluster within a replication rather than averaging out across estimands. Figure A4 shows the realized FDR and FWER. Every procedure in Figure A4 stays at or below the target throughout, whether it controls the FDR or the family-wise error rate. Even under this adversarial structure BH holds the realized FDR at the target rather than exceeding it: it stays at $q$ across every configuration, while under the other DGPs it is often below $q$.

# D Propositions and Proofs

In this section, we provide formal statements and proofs to support claims we make about the classification tests in the main text. To describe the critical values for the acceptance regions in magnitude and sign-and-magnitude tests, we use $z_{\text{dir}}$ and $z_{\text{eq}}$, which are related to $a$ (the lower limit of the upper acceptance region) and $b$ (the upper limit of the "negligible" acceptance region) by $a = \delta + z_{\text{dir}}\sigma$ and $b = \delta - z_{\text{eq}}\sigma$.

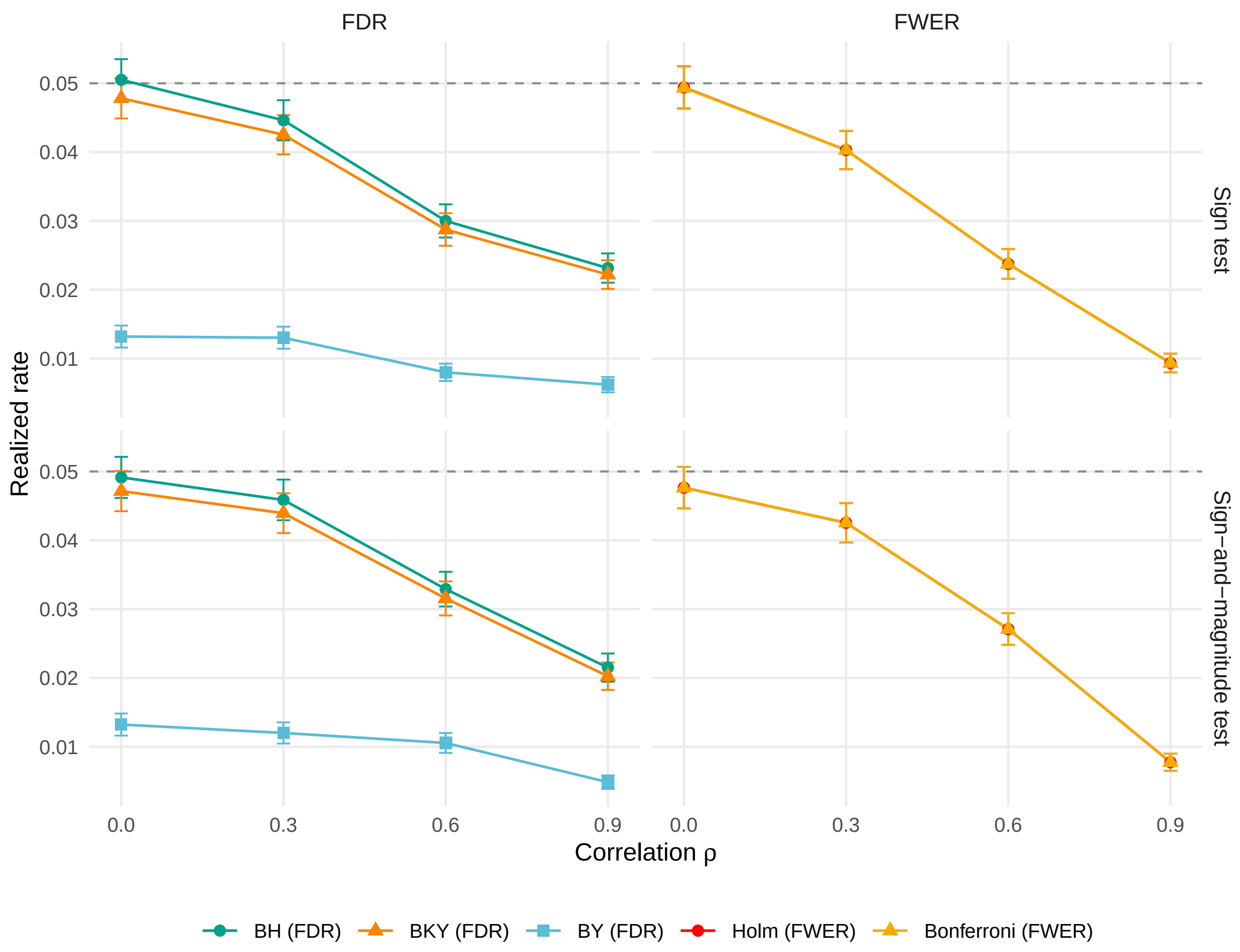

Figure A3: Realized FDR (left) and FWER (right) for the sign test (top) and the sign-and-magnitude test (bottom), against the correlation $\rho$ of the $K = 30$ estimates ($\sigma = 1$; 20,000 replications; bars indicate $\pm 2$ Monte Carlo standard errors). Each panel shows the worst-case configuration: every $\theta_k = 0$ for the sign test (each estimand at the class boundary) and every $\theta_k$ just below $\delta = 3$ for the sign-and-magnitude test. The dashed line marks the target $q = \alpha = 0.05$. Holm and Bonferroni have similar performance in these worst-case configurations so the lines are nearly overlapping.

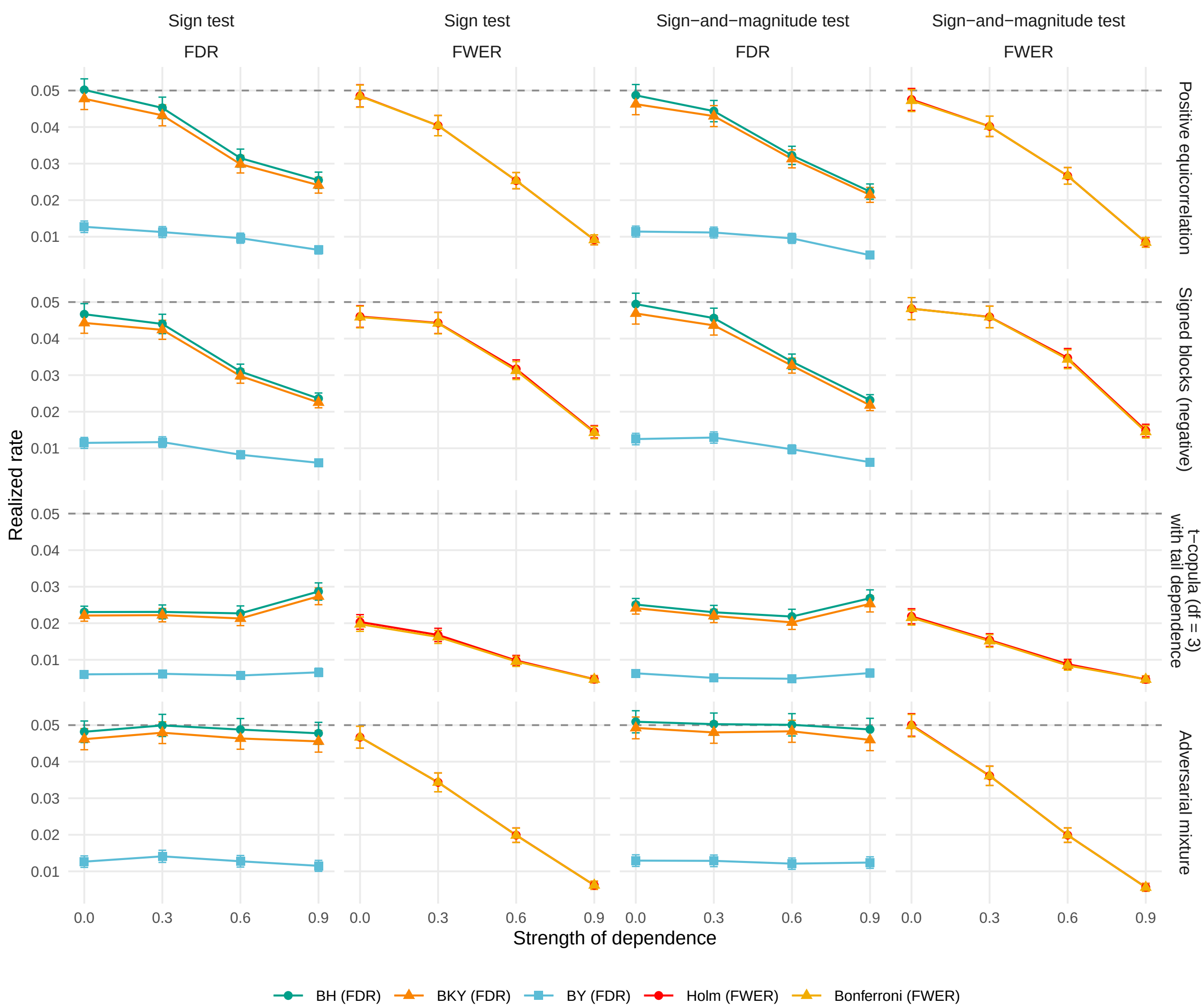

Figure A4: Realized FDR and FWER for the sign and sign-and-magnitude tests under four dependence structures (rows), against a parameter that controls the strength of dependence ($K = 30$; $\sigma = 1$; 20,000 replications; bars span $\pm 2$ Monte Carlo standard errors). Every structure has the same $N(\theta_k, 1)$ marginals; only the joint dependence varies. The rows are positive equicorrelation (correlation $+\rho$), signed blocks (correlation $+\rho$ within a block and $-\rho$ across blocks), a t-copula (df 3) with tail dependence, and a mixture in which all $K$ estimates share one shock with a probability equal to the dependence strength. Each test is run at the worst-case configuration (every $\theta_k = 0$ for the sign test; every $\theta_k$ just below $\delta = 3$ for the sign-and-magnitude test). The dashed line marks the target $q = \alpha = 0.05$. Holm and Bonferroni have similar performance in these worst-case configurations so the lines are nearly overlapping.

## D.1 Proposition (Location of Worst-Case Error)

**Proposition 1** (Location of Worst-Case Error). *Under a classification method with a single critical value $z$, the misclassification probability approaches its supremum as $\theta$ tends to a category boundary from inside the central region $C_0$. For a three-region partition with boundaries $\pm\delta$, the supremum is approached as $\theta \to \delta^-$ and $\theta \to -\delta^+$.*

*Proof.* We verify that the error reaches its maximum at the boundary $\theta = \delta$ (and by symmetry at $\theta = -\delta$) by checking three regions.

Fix any critical value $z > 0$ and let $Z = (\hat{\theta} - \theta)/\sigma \sim N(0, 1)$.

**(a)** For $\theta \in [0, \delta)$ (true category $C_0$), the two possible misclassifications assign $\hat{\theta}$ to the "positive and substantial" or "negative and substantial" category. The total error is

$$P(\text{error}) = [1 - \Phi(z + (\delta - \theta)/\sigma)] + \Phi(-z - (\delta + \theta)/\sigma).$$

Differentiating with respect to $\theta$ yields $(1/\sigma)[\phi(z+(\delta-\theta)/\sigma) - \phi(z+(\delta+\theta)/\sigma)]$. Both arguments exceed $z > 0$, and $z+(\delta-\theta)/\sigma < z+(\delta+\theta)/\sigma$ for $\theta > 0$, so $\phi$ evaluated at the first argument exceeds $\phi$ at the second (since $\phi$ is decreasing on $(0, \infty)$). The derivative is positive, so the error is increasing in $\theta$ on $[0, \delta)$, with the supremum approached as $\theta \to \delta^-$. The argument for $\theta \in (-\delta, 0]$ is symmetric.

**(b)** For $\theta \in [\delta, \infty)$ (true category $C_+$), let $u = \theta/\sigma \geq r$. The "negative and substantial" probability $P_- = \Phi(-r - z - u)$ is strictly decreasing in $u$ for $u > r$. When $z < r$ (the "negligible" acceptance region is non-empty), the "negligible"-misclassification probability is

$$P_0 = \Phi(r - z - u) - \Phi(-r + z - u),$$

with $dP/du = -\phi(r - z - u) + \phi(-r + z - u) = -\phi(r - z - u) + \phi(r - z + u)$

using $\phi$ symmetry. For $u > r > z > 0$, $(r - z + u) > (r - z - u)$ in absolute value, so $\phi(r-z+u) < \phi(r-z-u)$ and $dP/du < 0$. The total error therefore decreases strictly in $\theta$ for $\theta > \delta$. When $z \geq r$, the "negligible" region is empty and only the "negative and substantial" tail contributes to the error. That tail is also strictly decreasing in $u$. In either regime, the supremum on $[\delta, \infty)$ is attained at $\theta = \delta$ (i.e., $u = r$).

**(c)** For $\theta \in (-\infty, -\delta]$, the argument is symmetric to (b). □

**Corollary 1** (Boundary error of a classification rule with a single critical value)**.** *For a classification rule with a single critical value $z$ and a three-region partition with boundaries $\pm\delta$, the worst-case misclassification probability of a negligible estimand, approached as $\theta \to \delta^-$, equals $E(r, z) = [1 - \Phi(z)] + \Phi(-2r - z)$.*

*Proof.* By Proposition 1, the worst case occurs as $\theta \to \delta^-$ from inside $C_0$. In that limit the "positive and substantial" probability tends to $1 - \Phi(z)$ and the "negative and substantial" probability tends to $\Phi(-2r - z)$. The two events are mutually exclusive, so the limiting total error is their sum, $E(r, z) = [1 - \Phi(z)] + \Phi(-2r - z)$. □

## D.2 Proposition (Error Control)

**Proposition 2.** *The calibration method in the sign-and-magnitude classification test, which chooses $(z_{dir}, z_{eq})$ to satisfy Equation 3 and Equation 4, controls the overall misclassification rate at exactly $\alpha$: $\sup_\theta P(\text{misclassification}) = \alpha$.*

*Proof.* By symmetry, consider $\theta \geq 0$. Let $Z = (\hat{\theta} - \theta)/\sigma \sim N(0, 1)$.

**Case 1:** $\theta \in [0, \delta)$ **(true category is $C_0$).** The misclassifications are assignment to "positive and substantial" ($Z \geq z_{\text{dir}} + (\delta - \theta)/\sigma$) and assignment to "negative and substantial" ($Z \leq -z_{\text{dir}} - (\delta + \theta)/\sigma$). Both use $z_{\text{dir}}$ (which the calibration makes positive for $\alpha < 1/2$), so the analysis is identical to the proof of Proposition 1. The error is increasing in $\theta$ on $[0, \delta)$. The supremum is approached as $\theta \to \delta^-$, where the limit equals $\alpha$ by Equation 3.

**Case 2:** $\theta \in [\delta, \infty)$ **(true category is $C_+$).** Let $s = (\theta - \delta)/\sigma \geq 0$. The "negative and substantial" probability is $P_- = \Phi(-(s + 2r) - z_{\text{dir}})$. The "negligible"-misclassification probability is $P_0 = \max\{0, \Phi(-s-z_{\text{eq}})-\Phi(-(s+2r)+z_{\text{eq}})\}$. These events are mutually exclusive. Both $P_-$ and $P_0$ are strictly decreasing in $s$ for $s > 0$, so the supremum on $[\delta, \infty)$ is attained at $s = 0$ (i.e., $\theta = \delta$):

$$P_- + P_0\big|_{s=0} = \Phi(-2r - z_{\text{dir}}) + \max\{0, \Phi(-z_{\text{eq}}) - \Phi(-2r + z_{\text{eq}})\} = \alpha$$

by Equation 4. For $s > 0$, the error is strictly less than $\alpha$. $\square$

The next three propositions establish the optimality of the calibration in each classification test. All three follow the same pattern. For each class, the infimum of the correct-classification rate (over $\theta$ in that class) is attained at the class boundary. The size constraint sets this below $\alpha$. The maximin objective then forces equality for as many constraints as there are free parameters. The sign test has one parameter and one constraint, by symmetry. The magnitude test has two parameters and two independent constraints. The sign-and-magnitude test has two parameters and two corresponding constraints, through the wrong-sign probability $P_-$.

## D.3 Proposition (Optimality for the Sign Classification Test)

**Proposition 3.** *For the sign classification test with classes $C_- = (-\infty, 0]$, $C_+ = (0, \infty)$ and acceptance regions $A_- = (-\infty, -z\sigma]$, $A_+ = [z\sigma, \infty)$ parameterized by $z \geq 0$, the unique maximizer of*

$$W(z) := \min_j \inf_{\theta \in C_j} P_\theta(\hat{\theta} \in A_j)$$

*subject to the size constraint $\sup_\theta P_\theta(\text{misclassification}) \leq \alpha$ is $z^* = z_{1-\alpha}$, with $W^* = \alpha$.*

*Proof.* We first show that the symmetric one-parameter family is without loss. The problem is invariant under the reflection $\theta \mapsto -\theta$, which interchanges $C_+$ and $C_-$. The reflection of an acceptance pair $(A_-, A_+)$ is $(-A_+, -A_-)$, with the same $W$ and size. A maximin optimum can therefore be taken invariant, with $A_- = -A_+$. At fixed size the correct-classification probability on $C_+$ is maximized by an upper-tail region, because the Gaussian family has monotone likelihood ratio in $\hat{\theta}$. Hence $A_+ = [z\sigma, \infty)$ and, by invariance, $A_- = (-\infty, -z\sigma]$. It therefore suffices to optimize over $z \geq 0$.

For $\theta \in C_+$, $P_\theta(\hat{\theta} \in A_+) = 1 - \Phi(z - \theta/\sigma)$ is strictly increasing in $\theta$, so $\inf_{\theta \in C_+} P_\theta(\hat{\theta} \in A_+) = \lim_{\theta \downarrow 0} P_\theta(\hat{\theta} \in A_+) = 1 - \Phi(z)$. By symmetry under $\theta \mapsto -\theta$, $\inf_{\theta \in C_-} P_\theta(\hat{\theta} \in A_-) = P_{\theta=0}(\hat{\theta} \leq -z\sigma) = \Phi(-z) = 1 - \Phi(z)$. Therefore $W(z) = 1 - \Phi(z)$, continuous and strictly decreasing in $z$.

The size constraint binds at $\theta = 0$: over $\theta \in C_-$ the misclassification probability $P_\theta(\hat{\theta} \in A_+) = 1 - \Phi(z - \theta/\sigma)$ increases to its supremum $1 - \Phi(z)$ as $\theta \to 0^-$, so the constraint requires $1 - \Phi(z) \leq \alpha$, i.e., $z \geq z_{1-\alpha}$. Since $W$ is strictly decreasing, the unique maximizer is $z^* = z_{1-\alpha}$, giving $W^* = 1 - \Phi(z_{1-\alpha}) = \alpha$. □

## D.4 Proposition (Optimality for the Magnitude Classification Test)

**Proposition 4.** *For the magnitude classification test with classes $C_0 = (-\delta, \delta)$, $C_\pm = (-\infty, -\delta] \cup [\delta, \infty)$ and acceptance regions $A_0 = (-b, b)$, $A_\pm = (-\infty, -a] \cup [a, \infty)$ with $a, b \geq 0$, the unique maximizer of*

$$W(a, b) := \min_j \inf_{\theta \in C_j} P_\theta(\hat{\theta} \in A_j)$$

*subject to the size constraint $\sup_\theta P_\theta(\text{misclassification}) \leq \alpha$ is the pair $(a^*, b^*)$ uniquely determined by*

$$P_{\theta=\delta}(\hat{\theta} \in A_\pm) = \alpha \quad \text{and} \quad P_{\theta=\delta}(\hat{\theta} \in A_0) = \alpha.$$

*At the optimum, $W^* = \alpha$.*

*Proof.* We first reduce the search to a two-parameter family. The problem is invariant under the reflection $\theta \mapsto -\theta$. The classes are symmetric about $0$ ($C_0 = -C_0$ and $C_\pm = -C_\pm$), and if $\hat{\theta} \sim N(\theta, \sigma^2)$ then $-\hat{\theta} \sim N(-\theta, \sigma^2)$. The reflection of a rule with regions $(A_0, A_\pm)$ is the rule $(-A_0, -A_\pm)$, with the same $W$ and size. By this invariance, a maximin optimum can be taken symmetric, with $A_0 = -A_0$ and $A_\pm = -A_\pm$. Restricting to symmetric rules, the monotone likelihood ratio of the Gaussian family in $\hat{\theta}$ determines the shapes at fixed size: the correct-classification region for $C_0$ is a central interval and that for $C_\pm$ the complementary outer tails. This gives $A_0 = (-b, b)$ and $A_\pm = \{|\hat{\theta}| \geq a\}$, so it suffices to optimize over $(a, b)$.

For each class the correct-classification probability is monotone in $\theta$, attaining its infimum over the class at the boundary: $P_\theta(\hat{\theta} \in A_0) = \Phi((b-\theta)/\sigma) - \Phi((-b-\theta)/\sigma)$ is maximized at $\theta = 0$ and decreases as $|\theta| \to \delta$, while $P_\theta(\hat{\theta} \in A_\pm)$ increases as $\theta$ moves from $\delta$ into the tail. Hence each class infimum $\inf_{\theta \in C_j} P_\theta(\hat{\theta} \in A_j)$ equals the corresponding probability at the boundary $\theta = \pm\delta$. For the open class $C_0$ the boundary lies outside the class, so this infimum is the limit as $\theta \to \pm\delta$, equal to the boundary probability by continuity of $\Phi$. By symmetry evaluate at $\theta = \delta$:

$$\inf_{\theta \in C_0} P_\theta(\hat{\theta} \in A_0) = Q_0(b) := \Phi((b-\delta)/\sigma) - \Phi((-b-\delta)/\sigma),$$
$$\inf_{\theta \in C_\pm} P_\theta(\hat{\theta} \in A_\pm) = Q_\pm(a) := \left[1 - \Phi((a-\delta)/\sigma)\right] + \Phi((-a-\delta)/\sigma).$$

Thus $W(a, b) = \min\{Q_0(b),\, Q_\pm(a)\}$. The size constraint at $\theta = \delta$ gives $Q_\pm(a) \leq \alpha$ (the probability of misclassifying a borderline "negligible" $\theta$ as "substantial") and $Q_0(b) \leq \alpha$ (the probability of misclassifying a borderline "substantial" $\theta$ as "negligible"). These are two independent constraints, each involving only one parameter.

$Q_0$ is continuous and strictly increasing in $b$, with $Q_0(0) = 0$ and $Q_0 \to 1$ as $b \to \infty$. Likewise $Q_\pm$ is continuous and strictly decreasing in $a$, with $Q_\pm(0) = 1$ and $Q_\pm \to 0$ as $a \to \infty$. By the intermediate value theorem, $Q_0(b^*) = \alpha$ and $Q_\pm(a^*) = \alpha$ each have a unique solution. Feasibility requires $b \leq b^*$ and $a \geq a^*$, so

$$W(a, b) = \min\{Q_0(b),\, Q_\pm(a)\} \leq \min(\alpha, \alpha) = \alpha,$$

with equality iff $(a, b) = (a^*, b^*)$. For $\alpha < 1/2$, the identity $Q_0(x) + Q_\pm(x) = 1$ at any common argument $x$ gives $Q_\pm(b^*) = 1-\alpha > \alpha$, so by strict decrease of $Q_\pm$, $a^* > b^*$: the two acceptance regions are automatically disjoint at the optimum. □

## D.5 Proposition (Optimality for the Sign-and-Magnitude Classification Test)

**Proposition 5.** *Fix $\alpha < 1/2$. For a classification rule with pairwise-disjoint acceptance regions $A_+, A_-, A_0$, let*

$$W(A_+, A_-, A_0) := \min_j \inf_{\theta \in C_j} P_\theta(\hat{\theta} \in A_j)$$

*denote its worst-case power. Among all such rules whose size $\sup_\theta P_\theta(\textit{misclassification})$ is at most $\alpha$, $W$ is maximized uniquely by the rule with*

$$A_+ = [\delta + z^*_{dir}\sigma,\, \infty), \quad A_- = (-\infty,\, -\delta - z^*_{dir}\sigma], \quad A_0 = (-\delta + z^*_{eq}\sigma,\, \delta - z^*_{eq}\sigma),$$

*where $(z^*_{dir}, z^*_{eq})$ is the pair that solves Equation 3 and Equation 4 with equality.*

*Proof.* Consider the parametric rules with acceptance regions $A_+ = [\delta + z_{\text{dir}}\sigma, \infty)$, $A_- = (-\infty, -\delta - z_{\text{dir}}\sigma]$, and $A_0 = (-\delta + z_{\text{eq}}\sigma, \delta - z_{\text{eq}}\sigma)$, parameterized by the pair $(z_{\text{dir}}, z_{\text{eq}})$. These three regions are pairwise disjoint precisely when $z_{\text{eq}} \geq -z_{\text{dir}}$ and $z_{\text{dir}} > -r$.[24] Let $\Omega_\alpha$ be the set of parameter pairs $(z_{\text{dir}}, z_{\text{eq}})$ whose rule is pairwise disjoint and has size at most $\alpha$, and for such a pair write $W(z_{\text{dir}}, z_{\text{eq}})$ for the worst-case power of the corresponding rule.

We first show that a maximizer of $W$ over all size-$\alpha$ rules with pairwise-disjoint acceptance regions can be taken to be a parametric rule, so that it suffices to maximize $W(z_{\text{dir}}, z_{\text{eq}})$ over $\Omega_\alpha$. The problem is invariant under the reflection $\theta \mapsto -\theta$, which interchanges $C_+$ and $C_-$. The reflection of a rule $(A_+, A_-, A_0)$ is $(-A_-, -A_+, -A_0)$, with the same worst-case power and size. A maximizer can therefore be taken invariant, with $A_- = -A_+$ and $A_0 = -A_0$. At fixed size, the monotone likelihood ratio of the Gaussian family in $\hat{\theta}$ determines the shape of each region. The region that maximizes correct classification of $C_+$ is an upper half-line, and by invariance the region for $C_-$ is its reflection, a lower half-line. The region for $C_0$ is a central interval that is symmetric about $0$. THe classification thus uses a pairwise-disjoint parametric rule of size at most $\alpha$, so its parameter pair is in $\Omega_\alpha$. It therefore suffices to maximize $W(z_{\text{dir}}, z_{\text{eq}})$ over $\Omega_\alpha$, which we now do.

[24] The first inequality keeps $A_0$ disjoint from $A_+$ and $A_-$, and the second keeps $A_+$ disjoint from $A_-$. We thus rule out rules that assign an estimate to more than one class. Both hold at the calibrated optimum, where $z_{\text{dir}} > 0$.

We evaluate the three probabilities of having a definitive classification at the boundary $\theta = \delta$ (the probabilities at $\theta = -\delta$ follow by symmetry). Using $\hat{\theta} \sim N(\theta, \sigma^2)$ and $r = \delta/\sigma$,

$$
\begin{aligned}
P_+(z_{\text{dir}}) &:= P_{\theta=\delta}(\hat{\theta} \in A_+) = P_{\theta=\delta}(\hat{\theta} \geq \delta + z_{\text{dir}}\sigma) = 1 - \Phi(z_{\text{dir}}), \\
P_-(z_{\text{dir}}) &:= P_{\theta=\delta}(\hat{\theta} \in A_-) = P_{\theta=\delta}(\hat{\theta} \leq -\delta - z_{\text{dir}}\sigma) = \Phi(-2r - z_{\text{dir}}), \\
P_0(z_{\text{eq}}) &:= P_{\theta=\delta}(\hat{\theta} \in A_0) = \max\{0,\ \Phi(-z_{\text{eq}}) - \Phi(-2r + z_{\text{eq}})\}.
\end{aligned}
$$

The $\max\{0, \cdot\}$ in $P_0$ is for the corner case $z_{\text{eq}} \geq r$. Unlike the half-bounded acceptance regions $A_+$ and $A_-$ (which are never empty for finite $z_{\text{dir}}$), the "negligible" region $A_0 = (-\delta + z_{\text{eq}}\sigma,\ \delta - z_{\text{eq}}\sigma)$ is a bounded interval of width $2(\delta - z_{\text{eq}}\sigma)$ that is empty when $z_{\text{eq}}\sigma \geq \delta$ (i.e., $z_{\text{eq}} \geq r$). At that point $P_\theta(\hat{\theta} \in A_0) = 0$, while $\Phi(-z_{\text{eq}}) - \Phi(-2r + z_{\text{eq}})$ becomes negative. For $z_{\text{eq}} \in [0, r)$ the interval is non-empty, $\Phi(-z_{\text{eq}}) - \Phi(-2r + z_{\text{eq}}) > 0$, and $\max\{0,\ \Phi(-z_{\text{eq}}) - \Phi(-2r + z_{\text{eq}})\} = \Phi(-z_{\text{eq}}) - \Phi(-2r + z_{\text{eq}})$. We proceed in four steps.

**Step 1**. We show that $W$ reduces to $\min\{P_+(z_{\text{dir}}),\ P_0(z_{\text{eq}})\}$.

We compute $\inf_{\theta \in C_j} P_\theta(\hat{\theta} \in A_j)$ for each class $j \in \{+, -, 0\}$:

*Class* $C_+ = [\delta, \infty)$. For any $\theta \in C_+$,

$$\begin{aligned} P_\theta(\hat{\theta} \in A_+) &= P_\theta(\hat{\theta} \geq \delta + z_{\text{dir}}\sigma) \\ &= 1 - \Phi\Big(\frac{\delta + z_{\text{dir}}\sigma - \theta}{\sigma}\Big) \\ &= 1 - \Phi\big(z_{\text{dir}} - (\theta - \delta)/\sigma\big), \end{aligned}$$

which increases in $\theta$. As $\theta$ increases on $[\delta, \infty)$, $P_\theta(\hat{\theta} \in A_+)$ increases on $\theta \in [\delta, \infty)$. The infimum over $C_+$ is therefore attained at $\theta = \delta$, where $(\theta - \delta)/\sigma = 0$:

$$\inf_{\theta \in C_+} P_\theta(\hat{\theta} \in A_+) = 1 - \Phi(z_{\text{dir}}) = P_+(z_{\text{dir}}).$$

*Class* $C_- = (-\infty, -\delta]$. This case is the mirror image of $C_+$: the partition, $A_-$, and the sampling distribution all reflect consistently under $\theta \mapsto -\theta$. Concretely, for $\theta \in C_-$,

$$P_\theta(\hat{\theta} \in A_-) = P_\theta(\hat{\theta} \leq -\delta - z_{\text{dir}}\sigma) = \Phi\big(-z_{\text{dir}} + (-\delta - \theta)/\sigma\big).$$

Symmetrically, the infimum over $C_-$ is therefore attained at the right endpoint $\theta = -\delta$ (the right end of $C_-$, adjacent to $C_0$, where $(-\delta - \theta)/\sigma = 0$):

$$\inf_{\theta \in C_-} P_\theta(\hat{\theta} \in A_-) = \Phi(-z_{\text{dir}}) = 1 - \Phi(z_{\text{dir}}) = P_+(z_{\text{dir}}).$$

*Class* $C_0 = (-\delta, \delta)$. Set $u = \theta/\sigma$ (so $u \in (-r, r)$) and $c = r - z_{\text{eq}}$. The infimum below is over the open interval. By continuity of the normal CDF, it equals the limiting value at $u = \pm r$. Using $r\sigma = \delta$ from the definition of $r$, we can rewrite the acceptance region:

$$A_0 = (-\delta + z_{\text{eq}}\sigma,\, \delta - z_{\text{eq}}\sigma) = (-(r - z_{\text{eq}})\sigma,\, (r - z_{\text{eq}})\sigma) = (-c\sigma,\, c\sigma).$$

For $z_{\text{eq}} \geq -z_{\text{dir}}$, two regimes arise: if $z_{\text{eq}} > r$ then $c < 0$ and $A_0 = \emptyset$, so $P_\theta(\hat{\theta} \in A_0) = 0$ for all $\theta$ and the infimum equals $0 = P_0(z_{\text{eq}})$ (we use $\max\{0,\cdot\}$ in the definition of $P_0$ to handle this case). Henceforth in this step we treat the non-trivial regime $z_{\text{eq}} \le r$, so $c \ge 0$ and $A_0 = (-c\sigma, c\sigma)$ is non-empty. Standardizing $\hat{\theta} \sim N(\theta, \sigma^2)$ and substituting $u = \theta/\sigma$,

$$\begin{aligned} g(u) &:= P_\theta(\hat{\theta} \in A_0) \\ &= P_\theta(-c\sigma < \hat{\theta} < c\sigma) \\ &= \Phi((c\sigma - \theta)/\sigma) - \Phi((-c\sigma - \theta)/\sigma) \\ &= \Phi(c - u) - \Phi(-c - u). \end{aligned}$$

Let $\phi(\cdot)$ be the probability density function of the standard normal distribution. Differentiating $g(u)$, we have

$$g'(u) = -\phi(c - u) + \phi(c + u), \quad \text{using } \phi(-c - u) = \phi(c + u).$$

For $u > 0$ we have $|c+u| > |c-u|$: when $u \le c$, both are non-negative and $c+u > c-u$, and when $u > c$, $|c - u| = u - c < c + u = |c + u|$ since $c \ge 0$. Because $\phi$ is strictly decreasing in $|x|$, $\phi(c + u) < \phi(c - u)$, so $g'(u) < 0$ for $u > 0$. Also $g(u) = g(-u)$, so $g$ is symmetric about $u = 0$, strictly decreasing on $(0, r)$ and strictly increasing on $(-r, 0)$. The infimum of $g$ on the open interval $(-r, r)$ is therefore the limit at $u = \pm r$:

$$\begin{aligned} \inf_{\theta \in C_0} P_\theta(\hat{\theta} \in A_0) &= \lim_{u \to r^-} g(u) = g(r) \\ &= \Phi(-z_{\text{eq}}) - \Phi(-2r + z_{\text{eq}}) \\ &= P_0(z_{\text{eq}}). \end{aligned}$$

By symmetry the infimum over $C_-$ equals the infimum over $C_+$. Hence

$$W(z_{\text{dir}}, z_{\text{eq}}) = \min\{P_+(z_{\text{dir}}),\ P_0(z_{\text{eq}})\}.$$

**Step 2.** We show the size constraint reduces to two boundary inequalities.

By the proof of Proposition 2 (which generalizes Proposition 1 to the two-critical-value case), the worst-case misclassification probability within each class $C_j$ is attained at the boundary of $C_j$ nearest the adjacent class(es). Because the three acceptance regions $A_+, A_-, A_0$ are pairwise disjoint, the probability of any misclassification is the sum over incorrect classes.

*At $\theta = \delta$, approached from within $C_0$:* The correct class is "negligible" ($C_0$). The misclassifications are $C_+$ and $C_-$. Hence

$$\begin{aligned} P_{\theta=\delta}(\hat{\theta} \in A_+ \cup A_-) &= P_{\theta=\delta}(\hat{\theta} \in A_+) + P_{\theta=\delta}(\hat{\theta} \in A_-) \\ &= P_+(z_{\text{dir}}) + P_-(z_{\text{dir}}). \end{aligned}$$

*At $\theta = \delta$, approached from within $C_+$:* The correct class is "positive and substantial" ($C_+$). The misclassifications are $C_0$ and $C_-$. Hence

$$P_{\theta=\delta}(\hat{\theta} \in A_0 \cup A_-) = P_0(z_{\text{eq}}) + P_-(z_{\text{dir}}).$$

*At $\theta = -\delta$, approached from within $C_-$:* By symmetry, the misclassification probability equals $P_0(z_{\text{eq}}) + P_-(z_{\text{dir}})$, which is the same constraint as the $C_+$ boundary.

The size constraint $P_\theta(\hat{\theta} \in \bigcup_{k \neq j} A_k) \leq \alpha$ for all $j$ and all $\theta \in C_j$ is therefore equivalent to the conjunction of

$$\text{(A}'\text{)} \quad P_+(z_{\text{dir}}) + P_-(z_{\text{dir}}) \leq \alpha,$$

$$\text{(B}'\text{)} \quad P_0(z_{\text{eq}}) + P_-(z_{\text{dir}}) \leq \alpha,$$

that is, $\Omega_\alpha = \{(z_{\text{dir}}, z_{\text{eq}}) : z_{\text{eq}} \geq -z_{\text{dir}},\ z_{\text{dir}} > -r,\ (\text{A}'),\ (\text{B}')\}$, matching the definition above.

**Step 3.** We show that when the constraints A′ and B′ bind, the critical values $z^*_{\text{dir}}$ and $z^*_{\text{eq}}$ that satisfy the constraints are each unique.

**(i) Existence and uniqueness of $z^*_{\text{dir}}$.** The left-hand side of (A′) is

$$P_+(z_{\text{dir}}) + P_-(z_{\text{dir}}) = 1 - \Phi(z_{\text{dir}}) + \Phi(-2r - z_{\text{dir}}),$$

with derivative $-\phi(z_{\text{dir}}) - \phi(2r + z_{\text{dir}}) < 0$, so it is continuous and strictly decreasing in $z_{\text{dir}} \geq 0$. It equals $1/2 + \Phi(-2r) \in (1/2, 1)$ at $z_{\text{dir}} = 0$ and tends to $0$ as $z_{\text{dir}} \to \infty$. By the intermediate value theorem, there is a unique $z^*_{\text{dir}} > 0$ satisfying

$$P_+(z^*_{\text{dir}}) + P_-(z^*_{\text{dir}}) = \alpha. \tag{A}$$

Every feasible $z_{\text{dir}}$ (satisfying (A′)) satisfies $z_{\text{dir}} \geq z^*_{\text{dir}}$. Define $P^*_- := P_-(z^*_{\text{dir}})$ and $P^*_+ := P_+(z^*_{\text{dir}})$. From (A), $P^*_+ = \alpha - P^*_- > 0$.

**(ii) Existence and uniqueness of $z^*_{\text{eq}}$ given $z^*_{\text{dir}}$.** Setting $z_{\text{dir}} = z^*_{\text{dir}}$ in (B′) gives $P_0(z_{\text{eq}}) \leq \alpha - P^*_- = P^*_+$. The function $P_0$ is continuous on $[-z^*_{\text{dir}}, \infty)$, strictly decreasing on $[-z^*_{\text{dir}}, r]$ (derivative $-\phi(z_{\text{eq}}) - \phi(2r - z_{\text{eq}}) < 0$ on this interval, where $P_0(z_{\text{eq}}) = \Phi(-z_{\text{eq}}) - \Phi(-2r + z_{\text{eq}})$), and identically $0$ on $(r, \infty)$ (where $A_0$ is empty). To locate $z^*_{\text{eq}}$, we place $P^*_+$ in the interval between the endpoint values of $P_0$ on $[-z^*_{\text{dir}}, r]$. At the right endpoint $z_{\text{eq}} = r$, $P_0(r) = 0 < P^*_+$. At the left endpoint $z_{\text{eq}} = -z^*_{\text{dir}}$, the three acceptance regions $A_+, A_-, A_0$ become adjacent and partition $\mathbb{R}$ (at this specific boundary point only, not generically). Since $P_+(z^*_{\text{dir}}) + P_-(z^*_{\text{dir}}) + P_0(-z^*_{\text{dir}}) = 1$, constraint (A) gives $P_0(-z^*_{\text{dir}}) = 1 - \alpha > P^*_+$ (using $P^*_+ < \alpha < 1 - \alpha$ for $\alpha < 1/2$).

By the intermediate value theorem and strict monotonicity on $[-z^*_{\text{dir}}, r]$, there is a unique $z^*_{\text{eq}} \in (-z^*_{\text{dir}}, r)$ with

$$P_0(z^*_{\text{eq}}) = \alpha - P^*_- = P^*_+. \tag{B}$$

Set $P_0^* := P_0(z_{\text{eq}}^*) = P_+^*$. Uniqueness extends to the full half-line $z_{\text{eq}} \geq -z_{\text{dir}}^*$: on $(r, \infty)$, $P_0 = 0 \neq P_+^*$, so no other $z_{\text{eq}}$ satisfies (B). At $z_{\text{dir}} = z_{\text{dir}}^*$, (B$'$) holds iff $z_{\text{eq}} \geq z_{\text{eq}}^*$.

**Step 4.** We show that $(z_{\text{dir}}^*, z_{\text{eq}}^*)$ is the unique maximizer of $W$ in $\Omega_\alpha$.

From (A) and (B), at $(z_{\text{dir}}^*, z_{\text{eq}}^*)$, $P_+^* = P_0^* = \alpha - P_-^*$. Combined with Step 1's $W = \min\{P_+, P_0\}$,

$$
\begin{aligned}
W^* := W(z_{\text{dir}}^*, z_{\text{eq}}^*) &= \min\{P_+^*,\ P_0^*\} \\
&= P_+^* \\
&= \alpha - P_-^*.
\end{aligned}
$$

Let $(z_{\text{dir}}, z_{\text{eq}}) \in \Omega_\alpha$ be arbitrary. By Step 3, $z_{\text{dir}} \geq z_{\text{dir}}^*$ (otherwise (A$'$) is violated). Partition into three cases:

**Case (i): $z_{\text{dir}} > z_{\text{dir}}^*$.** Because $P_+(\cdot)$ is strictly decreasing, $P_+(z_{\text{dir}}) < P_+(z_{\text{dir}}^*) = P_+^*$. Since $W \leq P_+(z_{\text{dir}})$ (as $W$ is a minimum that includes $P_+$),

$$
W(z_{\text{dir}}, z_{\text{eq}}) \leq P_+(z_{\text{dir}}) < P_+^* = W^*.
$$

**Case (ii): $z_{\text{dir}} = z_{\text{dir}}^*$ and $z_{\text{eq}} > z_{\text{eq}}^*$.** Then $P_+(z_{\text{dir}}) = P_+^*$. By Step 3, $P_0$ is strictly decreasing on $[z_{\text{eq}}^*, r]$ and identically zero on $(r, \infty)$, so $P_0(z_{\text{eq}}) < P_0(z_{\text{eq}}^*) = P_0^* = P_+^*$ for any $z_{\text{eq}} > z_{\text{eq}}^*$. Hence

$$
\begin{aligned}
W(z_{\text{dir}}, z_{\text{eq}}) &= \min\{P_+^*,\ P_0(z_{\text{eq}})\} \\
&= P_0(z_{\text{eq}}) \\
&< P_+^* = W^*.
\end{aligned}
$$

**Case (iii): $z_{\text{dir}} = z_{\text{dir}}^*$ and $z_{\text{eq}} = z_{\text{eq}}^*$.** Then $W = W^*$.

The remaining possibilities, $z_{\text{dir}} < z_{\text{dir}}^*$ (infeasible by (A$'$)) and $z_{\text{dir}} = z_{\text{dir}}^*$ with $z_{\text{eq}} < z_{\text{eq}}^*$ (infeasible by (B$'$)), do not belong to $\Omega_\alpha$.

Therefore $W(z_{\text{dir}}, z_{\text{eq}}) < W^*$ for every $(z_{\text{dir}}, z_{\text{eq}}) \in \Omega_\alpha$ with $(z_{\text{dir}}, z_{\text{eq}}) \neq (z^*_{\text{dir}}, z^*_{\text{eq}})$, so $(z^*_{\text{dir}}, z^*_{\text{eq}})$ uniquely maximizes $W(z_{\text{dir}}, z_{\text{eq}})$ over $\Omega_\alpha$. By the reduction we established at the beginning of this proof, the rule with parameters $(z^*_{\text{dir}}, z^*_{\text{eq}})$ uniquely maximizes the worst-case power among all size-$\alpha$ rules with pairwise-disjoint acceptance regions. □

## D.6 Proposition (Asymptotic Validity with Estimated Standard Errors)

**Proposition 6.** *Assume the standard error estimator is consistent: $\hat{\sigma} \xrightarrow{p} \sigma$. The calibration method with estimated $\sigma$ then satisfies $\sup_\theta P_\theta(\textit{misclassification}) \xrightarrow{p} \alpha$.*

*Proof.* Write $\tilde{z}_d = z^*_{\text{dir}}(\hat{r})$ and $\tilde{z}_e = z^*_{\text{eq}}(\hat{r})$ for the procedure's (random) critical values, with corresponding acceptance regions $\tilde{A}_+, \tilde{A}_-, \tilde{A}_0$. We proceed in two steps: reduce the supremum over $\theta$ to the two boundary values at $\theta = \delta$, then apply the continuous mapping theorem.

**Step 1.** We reduce $\sup_\theta P_\theta(\text{misclassification})$ to the two misclassification probabilities at the boundary $\theta = \delta$.

Condition on $\hat{\sigma}$, so the acceptance regions $\tilde{A}_+, \tilde{A}_-, \tilde{A}_0$ are deterministic. The monotonicity argument in the proof of Proposition 2 relies only on the structural form of the regions (two symmetric outer half-lines and a symmetric central interval), not on the specific values of $(z_{\text{dir}}, z_{\text{eq}})$: the misclassification probability increases in $\theta$ on $[0, \delta)$ and decreases on $[\delta, \infty)$. The supremum over $\theta \geq 0$ is therefore the larger of the two boundary values,

$$
\begin{aligned}
M(\hat{\sigma}) &:= \max\{M_A(\hat{\sigma}),\, M_B(\hat{\sigma})\}, \\
\text{where } M_A(\hat{\sigma}) &:= P_{\theta \to \delta^-}(\hat{\theta} \in \tilde{A}_+ \cup \tilde{A}_- \mid \hat{\sigma}), \\
M_B(\hat{\sigma}) &:= P_{\theta = \delta}(\hat{\theta} \in \tilde{A}_- \cup \tilde{A}_0 \mid \hat{\sigma}).
\end{aligned}
$$

$M_A$ is the misclassification probability of a negligible estimand approaching the boundary from inside $C_0$ (misassigned to either substantial class) and $M_B$ that of a substantial estimand at the boundary (misassigned to the wrong substantial class or to negligible). At $\hat{\sigma} = \sigma$ both equal $\alpha$, by Constraints A and B. For $\hat{\sigma} \neq \sigma$ they differ, so the maximum $M(\hat{\sigma})$ is $M_A$ when $\hat{\sigma} < \sigma$ and $M_B$ when $\hat{\sigma} > \sigma$. The value at $\theta = -\delta$ is the same by symmetry. The remaining step shows both terms converge to $\alpha$.

**Step 2.** We show that $M(\hat{\sigma}) \xrightarrow{p} \alpha$ via the continuous mapping theorem.

Recall that we assume $\hat{\sigma} \xrightarrow{p} \sigma$. We also have the continuity of $\sigma \mapsto \delta/\sigma$ on $\sigma > 0$. Thus $\hat{r} \xrightarrow{p} r$. The calibration functions $z^*_{\text{dir}}$ and $z^*_{\text{eq}}$ are continuous in $r$ on the positive line, as they are smooth (by the implicit function theorem applied to Constraints A and B). We have $z^*_{\text{eq}}(r) < r$ for every $r > 0$ (see the proof for Proposition 5, Step 3), so the negligible region is always non-empty and no separate boundary case arises. The continuous mapping theorem gives $\tilde{z}_d \xrightarrow{p} z^*_{\text{dir}}(r)$ and $\tilde{z}_e \xrightarrow{p} z^*_{\text{eq}}(r)$.

Next, we compute the two boundary values. At $\theta = \delta$, $\hat{\theta} \sim N(\delta, \sigma^2)$. Standardizing and substituting the region forms,

$$
\begin{aligned}
M_A(\hat{\sigma}) &= \left[1 - \Phi(\tilde{z}_d \cdot \hat{\sigma}/\sigma)\right] + \Phi(-2r - \tilde{z}_d \cdot \hat{\sigma}/\sigma), \\
M_B(\hat{\sigma}) &= \Phi(-2r - \tilde{z}_d \cdot \hat{\sigma}/\sigma) + \max\left\{0,\ \Phi(-\tilde{z}_e \cdot \hat{\sigma}/\sigma) - \Phi(-2r + \tilde{z}_e \cdot \hat{\sigma}/\sigma)\right\}.
\end{aligned}
$$

Each right-hand side is a continuous function of $(\tilde{z}_d, \tilde{z}_e, \hat{\sigma}/\sigma)$. Applying the continuous mapping theorem with $\tilde{z}_d \xrightarrow{p} z^*_{\text{dir}}(r)$, $\tilde{z}_e \xrightarrow{p} z^*_{\text{eq}}(r)$, and $\hat{\sigma}/\sigma \xrightarrow{p} 1$,

$$
\begin{aligned}
M_A(\hat{\sigma}) &\xrightarrow{p} \left[1 - \Phi(z^*_{\text{dir}}(r))\right] + \Phi(-2r - z^*_{\text{dir}}(r)) = \alpha, \\
M_B(\hat{\sigma}) &\xrightarrow{p} \Phi(-2r - z^*_{\text{dir}}(r)) + \max\left\{0,\ \Phi(-z^*_{\text{eq}}(r)) - \Phi(-2r + z^*_{\text{eq}}(r))\right\} = \alpha,
\end{aligned}
$$

the first equality being Constraint A and the second Constraint B at the calibrated point $(z^*_{\text{dir}}(r), z^*_{\text{eq}}(r))$. Hence $M(\hat{\sigma}) = \max\{M_A(\hat{\sigma}), M_B(\hat{\sigma})\} \xrightarrow{p} \alpha$, and by Step 1 $\sup_\theta P_\theta(\text{misclassification}) \xrightarrow{p} \alpha$. $\square$

# E Comparison with Other Methods

This section compares our classification tests to other methods in two settings. Section E.1 compares the sign-and-magnitude test with four confidence-interval rules: the 90% CI, the 95% CI, the 90%/95% CIs (Goeman et al. 2010), and the Calibrated CI (a benchmark we construct). Section E.2 compares the magnitude classification test with TOST and the improved test of Berger and Hsu (1996).

## E.1 Variants of the sign-and-magnitude test

We compare our method to four confidence-interval rules via simulation. We set the target error level at $\alpha = 0.05$. Each confidence-interval rule assigns an estimand to a class when a confidence interval for $\theta$ lies entirely inside that class. Our method (classification testing) uses $z_{\text{dir}}(\hat{r})$ for directional claims and $z_{\text{eq}}(\hat{r})$ for equivalence claims, with both thresholds calibrated to the estimated precision ratio $\hat{r} = \delta/\hat{\sigma}$. The four alternative methods differ in which critical values they apply. The 90% CI uses the single critical value $z_{1-\alpha} = 1.645$ for every claim. The 95% CI uses the single critical value $z_{1-\alpha/2} = 1.96$ for every claim. The 90%/95% CIs of Goeman et al. (2010) use $z_{1-\alpha/2} = 1.96$ for directional claims and $z_{1-\alpha} = 1.645$ for equivalence claims. The Calibrated CI, a benchmark we construct, uses a single symmetric interval at coverage $1 - \gamma^*(r)$, where $\gamma^*(r)$ is the value of $\gamma$ at which the worst-case three-class misclassification rate equals $\alpha$. By Corollary 1, a symmetric interval of level $\gamma$ (critical value $z = z_{1-\gamma/2}$, so the near-side error $1 - \Phi(z)$ equals $\gamma/2$) has boundary error $\gamma/2 + \Phi(-2r - z_{1-\gamma/2})$, so $\gamma^*(r)$ solves $\gamma/2 + \Phi(-2r - z_{1-\gamma/2}) = \alpha$.

By construction, the 95% CI, the 90%/95% CIs, and the Calibrated CI should all have misclassification rates at or below $\alpha$, and our method should as well. The 90% CI does not have this property: its directional cutoff $z_{1-\alpha} = 1.645$ is too small to control the directional error, so its misclassification rate exceeds $\alpha$ near $\theta = \pm\delta$. For directional claims, both our method and the Calibrated CI use thresholds below $z_{1-\alpha/2}$ and therefore make (weakly) more classifications than the 95% CI or the 90%/95% CIs. For equivalence claims, our method uses $z_{\text{eq}} < z_{\text{dir}}$, which makes (weakly) more equivalence classifications than the other methods that control the error rate adequately. Our method thus (weakly) dominates the 95% CI, the 90%/95% CIs, and the Calibrated CI for all $r > 0$.

Each simulation draws $n = 500$ observations from $N(\theta, \sigma_0^2)$, computes $\bar{X}$ and $s/\sqrt{n}$, and classifies under all five methods. We repeat 20,000 times for each combination of $\theta$, $\sigma$, and $\delta$.

Figure A5 shows that four of the five methods control the misclassification rate at or below $\alpha = 0.05$: the 95% CI, the 90%/95% CIs, the Calibrated CI, and our method. The 95% CI is the most conservative. The 90% CI, on the other hand, is anticonservative: it exceeds $\alpha$ near $\theta = \pm\delta$, with a peak misclassification rate near $0.083$, because its single critical value $z_{1-\alpha} = 1.645$ fails to control the directional error at the 5% level. This shows using a 90% CI or 95% CI for classification can be either too restrictive or too permissive.

Figure A6 shows the probability of a correct classification for each of the four methods that control the error rate at or below the nominal level. Our method (weakly) dominates the alternatives, especially when the precision ratio $r = \delta/\sigma$ is small and the true value of $\theta$ does not have large magnitude. The 90%/95% CIs and the 95% CI share the directional cutoff $z_{1-\alpha/2}$, so they make the same directional classifications, but the 90%/95% CIs perform better than the 95% CI for negligible claims near $\theta = 0$. We do not include the 90% CI here because it does not guarantee control of the misclassification rate at the target level (see Figure A5).

## E.2 Magnitude Classification Testing vs Equivalence Testing via TOST and Berger-Hsu

We now compare the magnitude classification test (Section 5.2) to two alternatives: the standard TOST procedure and the improved test of Berger and Hsu (1996). Recall that the magnitude classification test defines $C_0 = (-\delta, \delta)$ and $C_\pm = (-\infty, -\delta] \cup [\delta, \infty)$, with acceptance region $A_0 = (-b, b)$ for "negligible" effects, where $b$ satisfies $P_{\theta=\delta}(\hat{\theta} \in (-b, b)) = \alpha$, and acceptance region $A_{\pm} = (-\infty, -a] \cup [a, \infty)$ for "substantial" effects, where $a$ satisfies $P_{\theta=\delta}(\hat{\theta} \in A_{\pm}) = \alpha$.

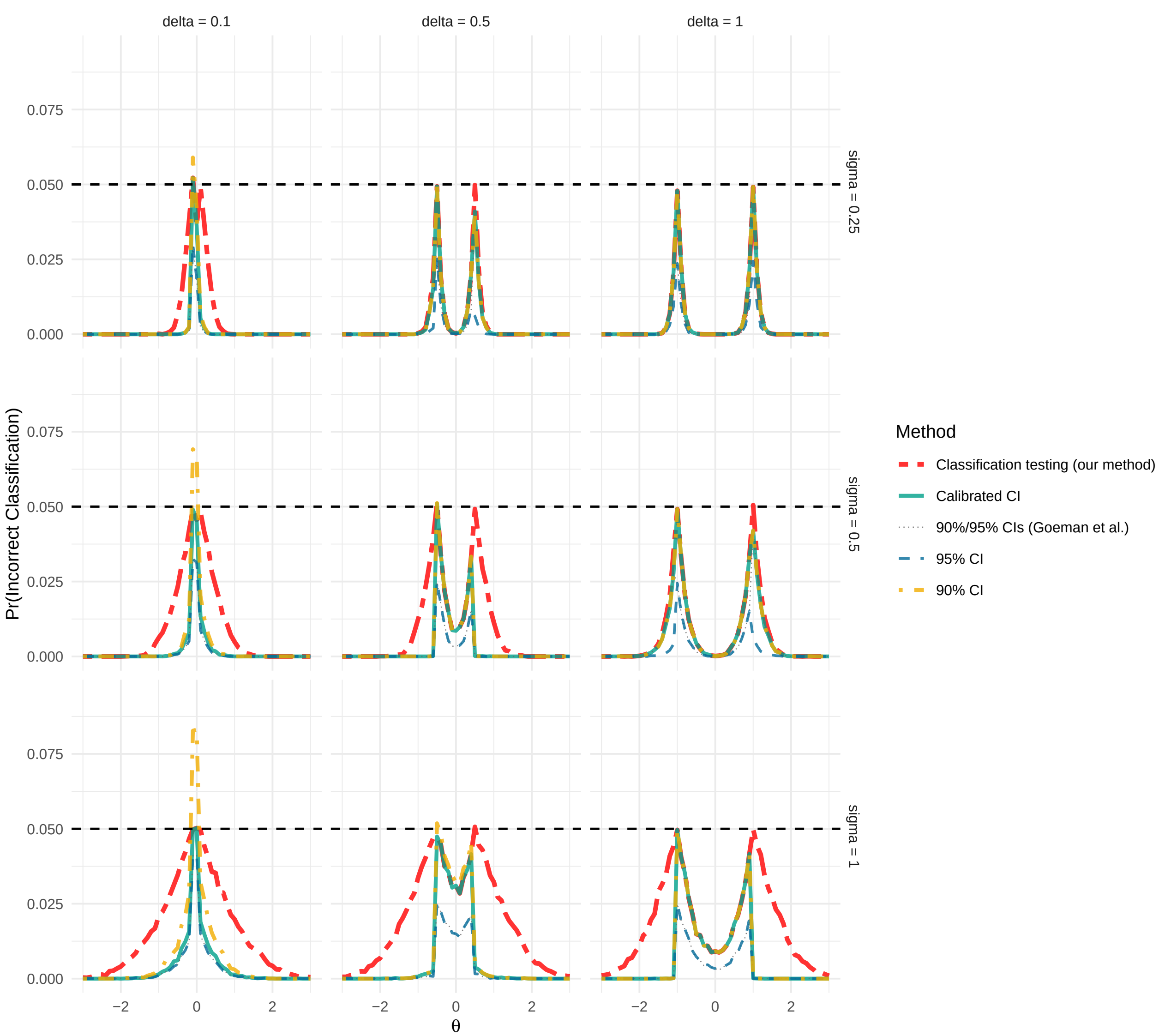

Figure A5: Overall misclassification rate across the five methods ($n = 500$, estimated $\sigma$, 20,000 replications). The 95% CI, the 90%/95% CIs (Goeman et al.), the Calibrated CI, and our method stay at or below $\alpha = 0.05$; the Calibrated CI and our method reach $\alpha$ at $\theta = \pm\delta$. The 90% CI exceeds $\alpha$ near $\theta = \pm\delta$ when the precision ratio is low (i.e., $\delta$ is small relative to $\sigma$).

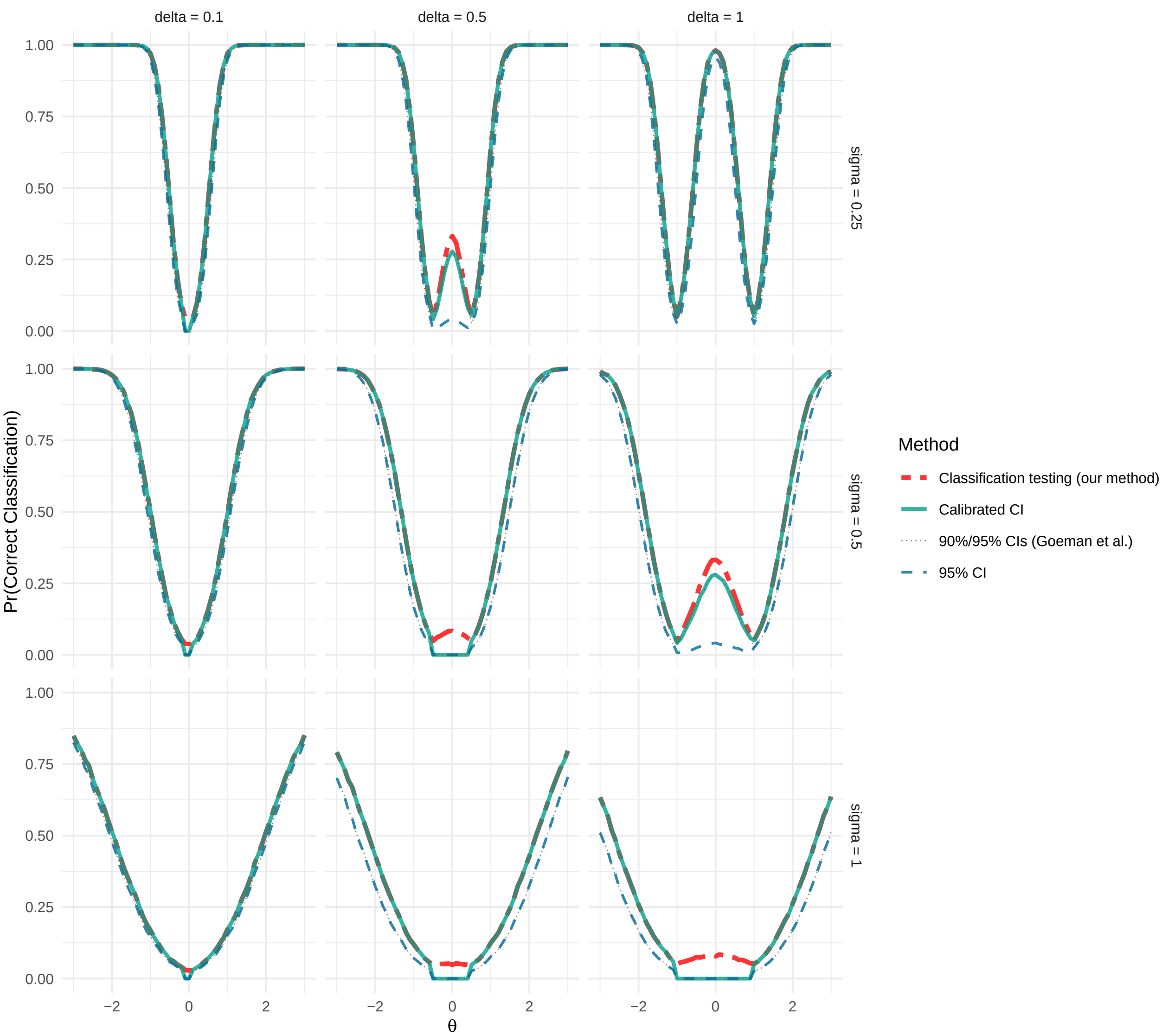

Figure A6: Probability of a correct classification for the four methods for sign-and-magnitude classification. Our method (weakly) dominates the other methods. We do not include the 90% CI here because it does not guarantee control of the misclassification rate at the target level (see Figure A5).

**Comparison to TOST.** The TOST acceptance region is $\{\hat{\theta} : |\hat{\theta}| < \delta - z_{1-\alpha}\sigma\}$, which exists only when $\sigma < \delta/z_{1-\alpha}$. For $\alpha = 0.05$, this means $\sigma < \delta/\Phi^{-1}(0.95) \approx 0.61\delta$. The classification test acceptance region, by contrast, always exists: it solves $\Phi((b-\delta)/\sigma) - \Phi((-b-\delta)/\sigma) = \alpha$ for $b > 0$, which has a solution for any $\sigma > 0$. The conservatism of TOST arises because each component one-sided test rejects for values of $\theta$ outside the equivalence region, where the null hypothesis of non-equivalence should not be rejected. The true error rate of TOST at the boundary $\theta = \delta$ is below $\alpha$, and more so as $\sigma$ increases. Our classification test calibrates the acceptance region to achieve exactly $\alpha$ at the boundary.

**Comparison to Berger and Hsu (1996).** The acceptance region for the "negligible" class in our magnitude classification test is the same as with the improved test of Berger and Hsu (1996) in the known-$\sigma$ case. Both approaches find the largest symmetric interval $[-b, b]$ such that $\sup_{|\theta| \geq \delta} P_\theta(\hat{\theta} \in [-b, b]) \leq \alpha$. The constraint binds at $\theta = \delta$, giving the same $b$.

The derivations are different. Berger and Hsu (1996) work top-down from the intersection-union test (IUT) framework, decomposing the null $H_0 : |\theta| \geq \delta$ into two component hypotheses and exploiting the dependence between the test statistics. Our approach works bottom-up from acceptance regions, solving directly for the boundary that uses the full $\alpha$ budget. In the known-$\sigma$ case, the IUT's "non-rectangular rejection region" reduces to a simple interval because the two test statistics are deterministic functions of $\hat{\theta}$.

The approaches differ when $\sigma$ is unknown. Berger and Hsu (1996) work with the $t$-distribution, where the unknown $\sigma$ cancels in the ratio $(\hat{\theta} - \theta)/\hat{\sigma}$. This yields exact finite-sample error control, but this finite-sample guarantee depends on a constant variance assumption. Our plug-in approach relies on normal approximation and replaces $\sigma$ with $\hat{\sigma}$.

For the sample sizes typical in political science experiments (hundreds to thousands), the difference between our plug-in approach and the exact Berger-Hsu test is negligible for equivalence testing. The practical advantage of our approach is simplicity: it requires only a normal CDF evaluation rather than non-central $t$ or $F$ distributions, and it extends naturally to quantities not expressed in standard deviation units.

## Appendix References